\documentclass[12pt]{article}           

\def\preprint{UTTG-15-97}       
\def\finished{April 16 1997}
\def\archive {hep-th/9704129}           

\long\def\abstract{                           
We explain the observation by Candelas and Font that the Dynkin diagrams
of nonabelian gauge groups occurring in type IIA and F-theory
can be read off from the polyhedron $\D^*$ that provides the toric
description of the Calabi--Yau manifold used for compacification.
We show how the intersection pattern of toric divisors corresponding
to the degeneration of elliptic fibers follows the $ADE$ classification of 
singularities and the Kodaira classification of degenerations.
We treat in detail the cases of elliptic $K3$ surfaces and $K3$
fibered threefolds where the fiber is again elliptic.
We also explain how even the occurrence of 
monodromy and non-simply laced groups in the latter case is visible in
the toric picture.
These methods also work in the fourfold case.
}

\let\Bbb=\mathbb        \def\CY{Calabi--Yau}

\def\beq{\BE} \def\eeq{\EE} \def\eeql{\EEL}  

\def\ipo{\hbox{\bf 0}}  

\def\fib{_{\rm fiber}} \def\bas{_{\rm base}} \def\Nt{{\tilde N}}

\input epsf

\usepackage{amssymb,latexsym}              \catcode`\"=\active \let"=\"

\def\ifundefined#1{\expandafter\ifx\csname#1\endcsname\relax}

\def\HS#1 {\hspace*{#1pt}} \def\VS#1 {\vspace*{#1pt}} \long\def\del#1\enddel{} 
\def\BC{\begin{center}}    
\def\EC{\end{center}}             

\newfont{\XHbf}{cmbx10 scaled 4800}     \newfont{\XH}{cmr10 scaled 4800}

\let\0=\over      \let\1=\vec      \def\2{{1\over2}}    \let\3=\ss
\let\4=\underline \let\5=\overline \let\6=\partial      \def\7#1{{#1}\llap{/}}
\def\8#1{{\textstyle{#1}}}         \def\9#1{{\ifmmode{\pmb{#1}}\else\bf#1\fi}}

          \def\({\left(}       \def\){\right)}

\def\EEL#1 {\label{#1}\EE}             
\def\BE {\begin{equation}}      \def\EE {\end{equation}}        
\def\BEA{\begin{eqnarray}}      \def\EEA{\end{eqnarray}} 

\def\pmb#1{\setbox0=\hbox{${#1}$}   \kern-.025em\copy0\kern-\wd0
      \kern.05em\copy0\kern-\wd0     \kern-.025em\raise.0433em\box0 }
\let\Bbb=\mathbb                                                        
\def\IR{{\Bbb R}} \def\IC{{\Bbb C}} \def\IP{{\Bbb P}} 
 \def\IN{{\Bbb N}}   

\let\and=\wedge

\let\bra=\langle        \let\ket=\rangle        \def\<#1\>{\bra #1 \ket}

\def\rel#1 #2{\buildrel #1 \over {#2}}  

\def\BP{\begin{picture}} \def\EP{\end{picture}}         
\def\putlab#1)#2#3{\put#1){\makebox(0,0)[#2]{\small #3}}}
\def\putlin#1,#2,#3,#4,#5){\put#1,#2){\line(#3,#4){#5}}} 
\def\putvec#1,#2,#3,#4,#5){\put#1,#2){\vector(#3,#4){#5}}}
\def\putc#1)#2{\put#1){\makebox(0,0)[c]{#2}}}
\def\Bezier#1#2)#3)#4){\qbezier#2)#3)#4)}       \thicklines

\def\subdef#1{\gdef\globalColor##1{##1}}      
      \subdef{Black}

\let\a=\alpha   \let\b=\beta    \let\g=\gamma      
         \let\th=\theta  
   \let\l=\lambda        
            \let\p=\pi            
\let\t=\tau                 
               \let\S=\Sigma 
             \let\D=\Delta

\def\plb#1 #2 {Phys. Lett. {\bf B#1} #2 }
\def\phr#1 #2 {Phys. Rep. {\bf  #1} #2 }        
\def\npb#1 #2 {Nucl. Phys. {\bf B#1} #2 }
\def\aph#1 #2 {Ann. Phys. {\bf #1} #2 }         
\def\jmp#1 #2 {J. Math. Phys. {\bf #1} #2 }
\def\jgp#1 #2 {J. Geom. Phys. {\bf #1} #2 }
\def\prd#1 #2 {Phys. Rev. {\bf D#1} #2 }
\def\prl#1 #2 {Phys. Rev. Lett. {\bf #1} #2 }
\def\rmp#1 #2 {Rev. Mod. Phys.  {\bf #1} #2 }
\def\zpc#1 {Z. Phys. {\bf #1C} }
\def\cmp#1 #2 {Commun. Math. Phys. {\bf #1} #2 }
\def\cqg#1 #2 {Class.Quant.Grav. {\bf #1} #2 }
\def\mpl#1 {Mod. Phys. Lett. {\bf A#1} }
\def\cpc#1 {Computer Phys. Commun. {\bf #1} }   
\def\ijmp#1 {Int. J. Mod. Phys. {\bf A#1} }
\def\ijmpC#1 {Int. J. Mod. Phys. {\bf C#1} }

\def\fnote#1#2{\begingroup\def\thefootnote{#1}\footnote{#2}
                \addtocounter{footnote}{-1}\endgroup}   

\def\offset#1#2{\def\xoff{#1}\def\yoff{#2}}     

\long\def\Fpar[#1](#2,#3)#4#5{\put(\xoff,\yoff){
                \put(#3,-#2){\makebox(0,0)[#1]{\parbox{#4mm}{#5}}}} }
\long\def\Fbox[#1](#2,#3)#4{\put(\xoff,\yoff){
                \put(#3,-#2){\makebox(0,0)[#1]{#4}}} }          \offset00

\newcounter{TRefNX} \let\OLDcite=\cite  \makeatletter
\def\makeTRefs#1{\@for  \NewTRef:=#1\do{\global\makeTRef{\NewTRef}}}
\def\makeTRef#1{\ifundefined{TRef#1}\stepcounter{TRefNX}%
\expandafter\xdef\csname TRef#1\endcsname{\theTRefNX}\fi}\makeatother
\def\NEWcite#1{\makeTRefs{#1}\OLDcite{#1}}  

\ifundefined{draftmode} {} \else        \let\cite=\NEWcite
\newcount\HOUR  \HOUR=\the\time \divide\HOUR by 60      \multiply \HOUR by 60 
\newcount\MIN   \MIN=\the\time  \advance\MIN by -\HOUR  \divide\HOUR by 60
\ifnum  \MIN>9  \def\printTIME{{\it\the\HOUR\,:\,\the\MIN}}
        \else   \def\printTIME{{\it\the\HOUR\,:\,0\the\MIN}} \fi 

\def\LLab#1{\BP(0,0)\unitlength=1mm\put(-12,.5){\makebox(0,0)[cr]{\small #1
        \rlap{$_{_{\makeatletter\csname TRef#1\endcsname\makeatother}}$}}}\EP}
\fi

\def\bye{\end{document}}                \long\def\new#1\endnew{{\bf #1}}
                                    
\begin{document}

\newfont{\XLbf}{cmbx10 scaled 2000}     \newfont{\XL}{cmr10 scaled 2000}
\newfont{\XLbfmath}{cmmi10 scaled 2000}

{\hfill \archive   \vskip -2pt \hfill\preprint }
\vskip 12mm
\centerline{\Large Enhanced Gauge Symmetry    }
\vskip 4mm
\centerline{\Large in Type II and F-Theory Compactifications: }
\vskip 4mm
\centerline{\Large Dynkin Diagrams From Polyhedra }
\begin{center} \vskip 7mm
        Eugene Perevalov\fnote{*}{e-mail: pereval@claude.ph.utexas.edu} 
\\[3mm]                       and
\\[3mm] Harald Skarke\fnote{\#}{e-mail: skarke@zerbina.ph.utexas.edu}
\\[8mm]{\it Theory Group, Department of Physics, University of Texas at 
        Austin\\
        Austin, TX 78712, USA}
        
\vfill                   \end{center}    \abstract

\vfill \noindent  \finished \vspace*{7mm}
\thispagestyle{empty} \newpage
\pagestyle{plain} 

\newpage
\setcounter{page}{1}
  
\section{Introduction}

Dualities and nonperturbative phenomena in string theory have been a subject
of extensive study for the last two years. Beginning with the work of refs.
\cite{HT,W95,Str} it has been realized that compactifications of string
theories on singular spaces lead to extra massless degrees of freedom 
and hence to a variety of possibilities for new connections between 
apparently different string theories. In particular, in \cite{W95} it was 
conjectured for the first time that if a type IIA string is compactified  
on a $K3$ surface with an orbifold singularity then the resulting theory 
can exhibit a simply-laced nonabelian gauge group of the type which matches
exactly the $ADE$ singularity type of the $K3$ surface in question. 
Consequently, similar statements were made about type II strings compactified
on \CY\ threefolds. Namely, in \cite{Str} it was shown that a conifold
type singularity leads to the appearance of massless black hole 
hypermultiplets in the low energy theory, and it was anticipated in 
\cite{Asp:dp,BSV,Asp:cy} and shown in \cite{KMP} that a curve of singularities
brings about the occurrence of an enhanced gauge symmetry. 
The latter fact, among other things, makes possible the duality
between heterotic strings on $K3\times T^2$ and type IIA on a \CY\ threefold
leading to $N=2$ supersymmetry in four dimensions, the  study of which was
pioneered in \cite{KV} and continued in numerous articles. 

On the type II side of this duality, the enhanced gauge symmetry appears
nonperturbatively and is due to the fact that the compactification manifold
is singular. It has to be a $K3$ fibration \cite{KLM,AL}, and if we are 
interested 
in constructing duals to perturbative heterotic vacua, the singularity
in question is an orbifold singularity of a generic fiber \cite{Asp:cy}.
The singularity type (according to $ADE$ classification) is nothing else
than the type of the gauge group appearing in the low energy theory.
Going to the Coulomb branch of the latter means resolving the singularities
by means of consecutive blow-ups. The intersection pattern of rational
curves introduced in the process reproduces the Dynkin diagram of the 
gauge group.  

Switching off Wilson lines on $T^2$ and making the torus big, this duality can 
be lifted to six dimensions. The \CY\ manifold then must be an elliptic 
fibration and the type IIA string becomes F-theory on that \CY\ \cite{F}.
F-theory provides a powerful tool for constructing duals to heterotic vacua
\cite{MVI,MVII}.
The gauge groups appearing in the compactified theory are due to the 
degeneration of the elliptic fiber over a codimension one 
locus in the base.
A detailed dictionary between such geometric data and physics was given
in \cite{six}. 

The authors of \cite{CF} studied the Heterotic/Type II duality in four
dimensions by means of toric geometry.
They constructed reflexive polyhedra describing the (resolved version of)
the singular \CY\ threefolds for the type II side. They observed a regular
structure present in the polyhedra. In particular, the generic $K3$ fiber 
could be identified as a subpolyhedron and the Dynkin diagrams of the gauge
groups present in four dimensions (including non-simply laced ones) were 
visible directly in the subpolyhedron. Later in \cite{BS,BS1,mon} it was 
further
demonstrated that the methods of toric geometry allow one to easily read
off the essential information from the corresponding polyhedra and construct 
\CY\ manifolds relevant to different aspects of string dualities. 
The relation between enhanced gauge symmetry and toric diagrams was also 
noted in \cite{KM,KKV}.

The purpose of the present article is to explain the appearance of Dynkin 
diagrams in the examples of \cite{CF} as well as give a dictionary between
the structure of the polyhedra and geometry (and hence physics) which 
is used in \cite{mon} to find the spectra of F-theory compactified on \CY\
threefolds with large numbers of K\"{a}hler class parameters.

The paper is organized as follows. In section 2, we give some background
information about divisors and intersection theory on toric varieties
and the construction of Calabi--Yau hypersurfaces.
In section 3, we show how to calculate the Picard lattice of a toric $K3$ 
manifold,
or in other words, how to determine the orbifold singularity type which results
upon blowing down the rational curves in the Picard lattice. In section 4, we
specialize to the case of an elliptic $K3$ and show that the Dynkin diagrams
of \cite{CF} appear in a natural way. Section 5 is devoted to the study of a
\CY\ threefold case. In particular we demonstrate the appearance of Dynkin 
diagrams of non-simply laced groups. Finally, section 6 summarizes our results
and contains a brief discussion of more complicated cases not covered in this
work.

\section{Toric Preliminaries}

\subsection{Divisors in Toric Varieties and their Intersections }

In this section we will give a brief review of the facts concerning
intersection theory on toric varieties which are relevant for the following
discussion.
We assume the reader's familiarity with basic notions of toric geometry,
such as the definitions of cones and fans 
(see, e.g., \cite{FU93} or \cite{OD88}).
We use standard notation, denoting the dual lattices by $M$ and $N$, their 
real extensions by $M_\IR$ and $N_\IR$, and the fan in $N_\IR$ by $\S$.

A fan $\S$ in $N_\IR$ defines a toric variety denoted by $X(\S)$. 
Let $\tau$ be any $k$-dimensional cone in $\S$ and
 $N_{\tau}$ be the sublattice of $N$ generated by $\tau \cap N$ and
\beq
N(\tau)=N/N_{\tau}.
\eeq
\del
The star of a cone can be defined abstractedly as the set of cones in $\S$
that contain $\tau$ as a face. Such cones $\sigma$ are determined by their
images in $N(\tau)$ which form a fan in $N(\tau)$ denoted by Star$(\tau)$. 
\enddel
The images in $N(\tau)$ of the cones in $\S$
that contain $\tau$ as a face form a fan in $N(\tau)$ denoted by Star$(\tau)$.
Then
\beq
V(\tau)=X({\rm Star}(\tau))
\eeq
is an $(n-k)$-dimensional closed subvariety of $X(\S)$. $V(\t)$ is called 
the orbit closure. (A toric variety $X(\S)$ is a disjoint union of orbits
$O_{\t}$ of the torus action, one such orbit corresponding to each cone $\t$
in $\S$.)


The Chow group $A_{n-k}$ on an arbitrary toric variety $X(\S)$ is generated 
by the classes of the orbit closures $V(\t)$ of the cones $\t$ of dimension 
$k$. In particular, each edge (one-dimensional cone) $\t_i$, generated by a
unique lattice vector $v_i$, 
gives rise to a T-Weil divisor
\beq 
D_i=V(\t_i).
\eeq 
Suppose $D$ is a Cartier divisor on a variety $X$ and $V$ is an irreducible
subvariety of $X$ which $D$ meets properly. In this case we can define an 
intersection cycle $D\cdot V$ by restricting $D$ to $V$, determining a Cartier
divisor $D\vert_{V}$ on $V$, and taking the Weil divisor of this Cartier 
divisor: $D\cdot V=[D\vert_V]$. Now, take $X$ to be a toric variety, $D=\sum
a_iD_i$ a T-Cartier divisor, and $V=V(\t)$. In this case we obtain
\beq
D\cdot V(\t)=\sum b_{\g}V(\g),
\eeq
where the sum is over all cones $\g$ containing $\t$ with dim$(\g)={\rm dim}
(\t)+1$, and $b_{\g}$ are integers computed in the following way. Suppose
$\g$ is spanned by $\t$ and a set of minimal edge vectors $v_i$. Let $e$ be
the generator of the one-dimensional lattice $N_{\g}/N_{\t}$ such that the 
image of each $v_i$ in $N_{\g}/N_{\t}$ is $s_i\cdot e$ with $s_i$ integers.
Then $b_{\g}$ is given by the formula
\beq
b_{\g}={a_i\over s_i},
\eeq
all $i$'s giving the same result.
If $X$ is nonsingular, then there is only one $i$ and $s_i=1$, so $b_{\g}$
is the coefficient of $D_i$ in $D$. In this case, $D_k$ is a Cartier divisor,
and 
\beq
 D_k\cdot V(\t)=\left\{ \begin{array}{ll}
                           V(\g)  &\mbox{if $\t$ and $v_k$ span a cone $\g$} \\
                       0 & \mbox{if $\t$ and $v_k$ don't span a cone in $\S$}
                         \end{array}
                 \right. 
\eeql{dkv}
We can now use (\ref{dkv}) one more time to obtain a formula for the triple
intersection $D_j\cdot D_k\cdot V(\t)$ and so on. In particular, when
$\t$ itself is a one-dimensional cone we obtain the following formula for 
the intersection of Cartier (Weil) divisors on a nonsingular toric variety
\beq
 D_{k_1}\cdot D_{k_2}\cdot \ldots \cdot D_{k_m}=\left\{
   \begin{array}{ll} 
        V(\g) &\mbox{if $v_{k_i}$, $i=1,\ldots m$, span a cone $\g$} \\
        0     &\mbox{if $v_{k_i}$, $i=1,\ldots m$, don't span a cone in $\S$}
     \end{array}
    \right.
\eeql{dkdk}
If we are interested in the intersection of $n$ divisors, i.e. $m=n$ in
(\ref{dkdk}), we arrive at the intersection number. Namely, provided the
$v_{k_i}$ in (\ref{dkdk}) span an $n$-dimensional cone $\g$,  
$D_{k_1}\cdot D_{k_2}\cdot \ldots \cdot D_{k_m}=X({\rm Star}(\g))$,
where ${\rm Star}(\g)$ is simply a toric variety of dimension zero, that is
a point. Thus, we finally obtain the formula for intersection number which
will be of most use for us in what follows.
\beq
D_{k_1}\cdot D_{k_2}\cdot \ldots \cdot D_{k_n}=\left\{
   \begin{array}{ll} 
        1 &\mbox{if $v_{k_i}$, $i=1,\ldots n$, span an $n$-dimensional 
       cone in $\S$} \\
        0     &\mbox{if $v_{k_i}$, $i=1,\ldots n$, don't span a cone in $\S$}
     \end{array}
    \right.
\eeql{num}

Not all of $D_k$'s are linearly independent though. There are certain 
relations between them which can be described as follows. Let $m\in M$.
Then it can be shown that
\beq
[{\rm div}(f)]=\sum_i \<m,v_i\>D_i,
\eeq
where $f$ is a certain nonzero rational function
and the sum extends over all one-dimensional cones in $\S$.
Thus div$(f)$ is a
principal divisor, i.e. the sum in the r.h.s. of (\ref{lrel}) is linearly
equivalent to zero:
\beq
\sum_i \<m,v_i\>D_i ~\sim ~0~~~~\forall~ m\in M.
\eeql{lrel}
It is obvious that we can write $n$ independent relations
of the form (\ref{lrel}), and hence, in general rank(Pic($X))\leq d_1-n$,
where $d_1$ is the number of one-dimensional cones in $\S$. If $X$ is 
nonsingular (in fact, it is sufficient to require that each cone in $\S$ be 
simplicial), the above inequality becomes an equality:
\beq
{\rm rank}({\rm Pic}(X))=d_1-n.
\eeql{picard}

There is a convenient way of describing toric varieties which has been 
introduced in \cite{cox} and will be used in what follows. Namely, to each
one-dimensional cone in $\S$ with primitive generator $v_k$ we can assign
a homogeneous coordinate $z_k$, $k=1,\ldots N$. Then we remove the set    
$Z_{\S}$ from the resulting $\IC^N$, where
\beq
Z_\S=\bigcup_I\{(z_1,\ldots,z_N):\; z_i=0\;\forall i\in I\}
\eeql{exset}
with the union taken over all sets 
$I\subseteq \{1,\cdots,N\}$ for which 
$\{v_i:\;i\in I\}$ does not belong to a cone in $\S$.

Then the toric variety $X(\S)$ is given by the quotient of 
$\IC^N\setminus Z_\S$ by $(\IC^*)^{N-n}$ 
(times, perhaps, a finite abelian group) whose action is given by
\beq
(z_1,\ldots,z_N)\sim (\l^{w^1_j}z_1,\ldots,\l^{w^N_j}z_N)~~~~~
{\rm if}~~~~\sum_k w^k_j v_k=0 
\eeql{er}
where $N-n$ of such linear relations are independent, and we can limit
ourselves by $j=1,\ldots ,N-n$.

In such a description the divisors $D_k$ are given simply by $z_k=0$ and
it is very easy to see, for example, that in the case when two one-dimensional
cones do not belong to a higher-dimensional cone in $\S$  the corresponding 
divisors do not intersect in $X(\S)$. Indeed, in this case we simply are not 
allowed to set both $z$'s to zero simultaneously (the resulting set falls into
$Z_\S$ as we can see from (\ref{exset}) and hence does not belong to our
variety). 

\subsection{Hypersurfaces of Vanishing First Chern class}

We will be dealing with \CY\ manifolds given as hypersurfaces in toric 
varieties \cite{ba93}. 
To such a manifold there correspond a reflexive polyhedron 
$\D\subset M_\IR$ and its dual $\D^*$.
The dual of any set $S\subset M_\IR$ is given by the set 
\beq S^*=\{y\in N_\IR:~\<x,y\>\ge -1\}, \eeq
with an analogous definition for duals of sets in $N_\IR$.
A reflexive polyhedron $\D$ is a lattice polyhedron (i.e., a polyhedron
whose vertices are lattice points) containing the lattice origin $\ipo$
such that its dual $\D^*$ is again a lattice polyhedron.
Then $(\D^*)^*=\D$, and in any lattice basis the coordinates of the
vertices of $\D^*$ are just the coefficients of the equations for the
bounding hyperplanes of $\D$, with the r.h.s. normalized to $-1$.

The fan $\S$ defining the ambient toric variety consists of cones that
are determined by some triangulation of $\D^*$, which we will always
assume to be maximal. 
In particular, the one-dimensional cones of $\S$ correspond to
the lattice points (except $\ipo$) of $\D^*$.
If we want to construct a hypersurface of trivial canonical class,
we must choose it as the zero locus of some section of an appropriate 
line bundle.
This line bundle is determined by a polynomial with monomials that
correspond to points in $\D$ in such a way that 
each lattice point $m\in\D$ gives rise to a monomial in the $z_i$'s
with exponents $\<m,v_i\>+1$.
The resulting polynomial is quasihomogeneous w.r.t. the relations (\ref{er}), 
with transformation properties that
correspond precisely to those of the monomial $\prod_{i=1}^Nz_i$.
Thus, viewed as a divisor in the ambient space, the Calabi--Yau
hypersurface is linearly equivalent to $\sum_{i=1}^ND_i$.

Provided the complex dimension of the manifold is greater than one, the 
rank of its Picard group can be calculated from the combinatorial data on
the polyhedron as follows. 
If every divisor of $X(\S)$ intersected the Calabi--Yau hypersurface precisely
once, then the Picard number of the hypersurface would be determined by 
formula (\ref{picard}) with $d_1=l(\D^*)-1$, where $l(\D^*)$ is 
the number of lattice points in 
$\D^*$ and the subtraction of 1 reflects that fact that the origin does not
correspond to a one-dimensional cone. From 
this number we have to subtract the number of divisors in $X(\S)$
that don't intersect the hypersurface at all; it is known (and we will 
rederive this fact for the $K3$ case) that these are precisely the 
divisors corresponding to points interior to facets (codimension one faces)
of $\D^*$.
Finally we have to add a correction term for the cases where a single divisor 
in $X(\S)$ leads to more than one divisor in the hypersurface.
Thus we arrive at
\beq
{\rm rank}({\rm Pic}(X_{\rm CY}))=
l(\D^{\ast})-\sum_{{\rm codim}(\th^*)=1}
l^{\prime}(\th^*)+\sum_{{\rm codim}(\th^*)=2}
l^{\prime}(\th^*)l^{\prime}(\theta)-(n+1),
\label{pic}
\eeq
where $\theta^*$ and $\theta$ are dual faces of $\D^*$ and $\D$,
respectively,
and $l^{\prime}(\theta)$ denotes the number 
of points in the interior of the face $\theta$. The first two terms in 
(\ref{pic}) count the number of points which are not interior to codimension
one faces. We will call these points relevant. The third term in (\ref{pic})
is the correction term.
\del
, it counts the number of divisor classes in $X$ which
are not represented torically. Since our task is to make statements about
manifolds given toric information 
\enddel
To keep the discussion manageable, we will sometimes assume in the following 
that the correction term vanishes.
This is not a severe restriction because in many cases it is possible to
pass from a toric description that requires this term to one that does not
\cite{ca95,CF}.
It will play an important role, however, when we discuss monodromies
and non-simply laced gauge groups in the context of $K3$
fibered Calabi--Yau threefolds.

\section{The Picard Lattice of a Toric $K3$ Surface} 

In this section we will analyze in detail the intersection pattern of
divisors in a $K3$ surface which is described by a three- dimensional 
reflexive 
polyhedron $\D^*$.
Our discussion will be valid whether or not the correction term in
(\ref{pic}) vanishes; in the latter case we will find that 
some of the divisors corresponding to points in $\D^*$ must be reducible .
The reader who is not interested in all the technical details of the 
calculations of intersection numbers may wish to jump to the summary
at the end of this section.

\begin{figure}[htb]
\epsfxsize=2in
\hfil\epsfbox{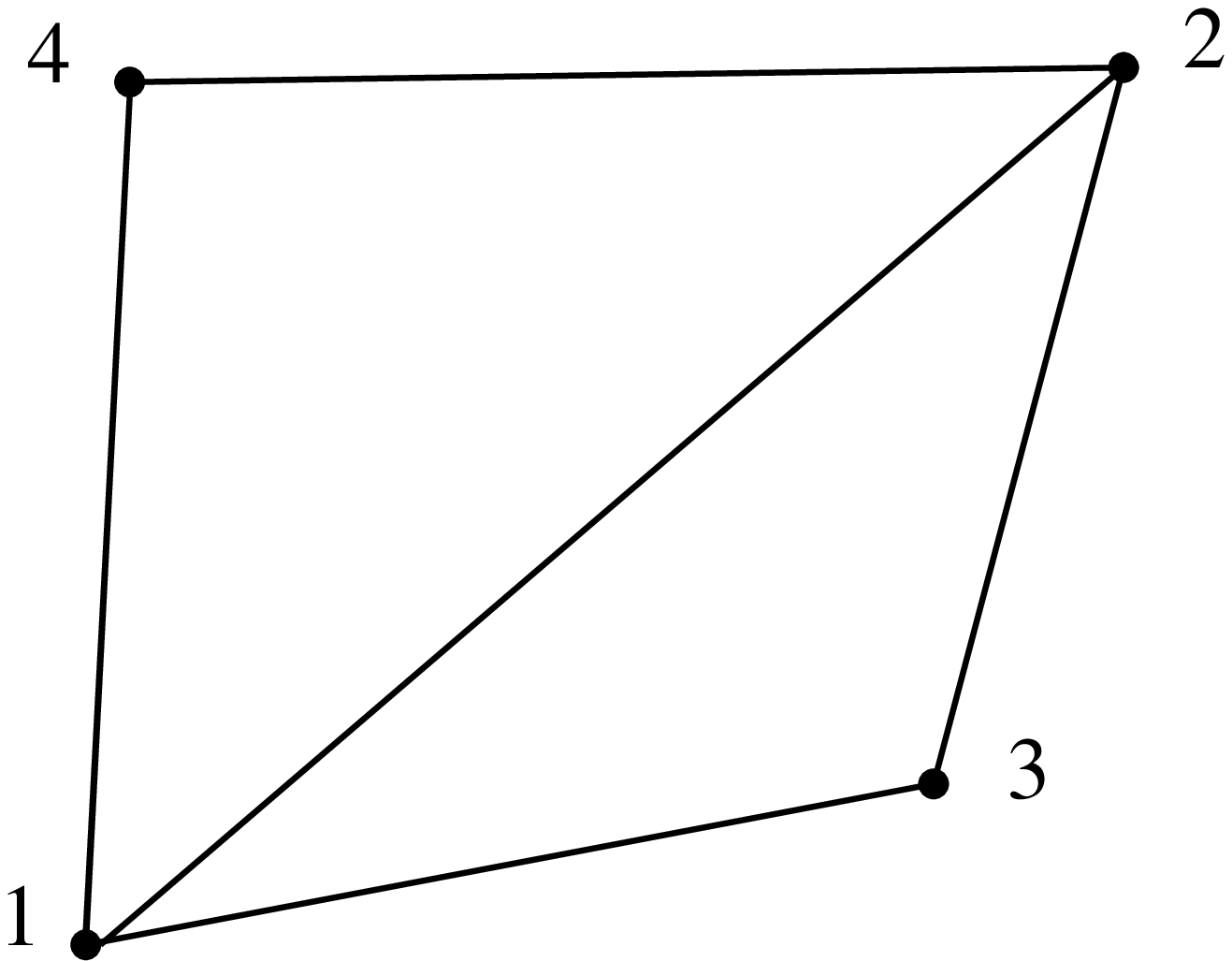}\hfil\epsfxsize=2in
\hfil\epsfbox{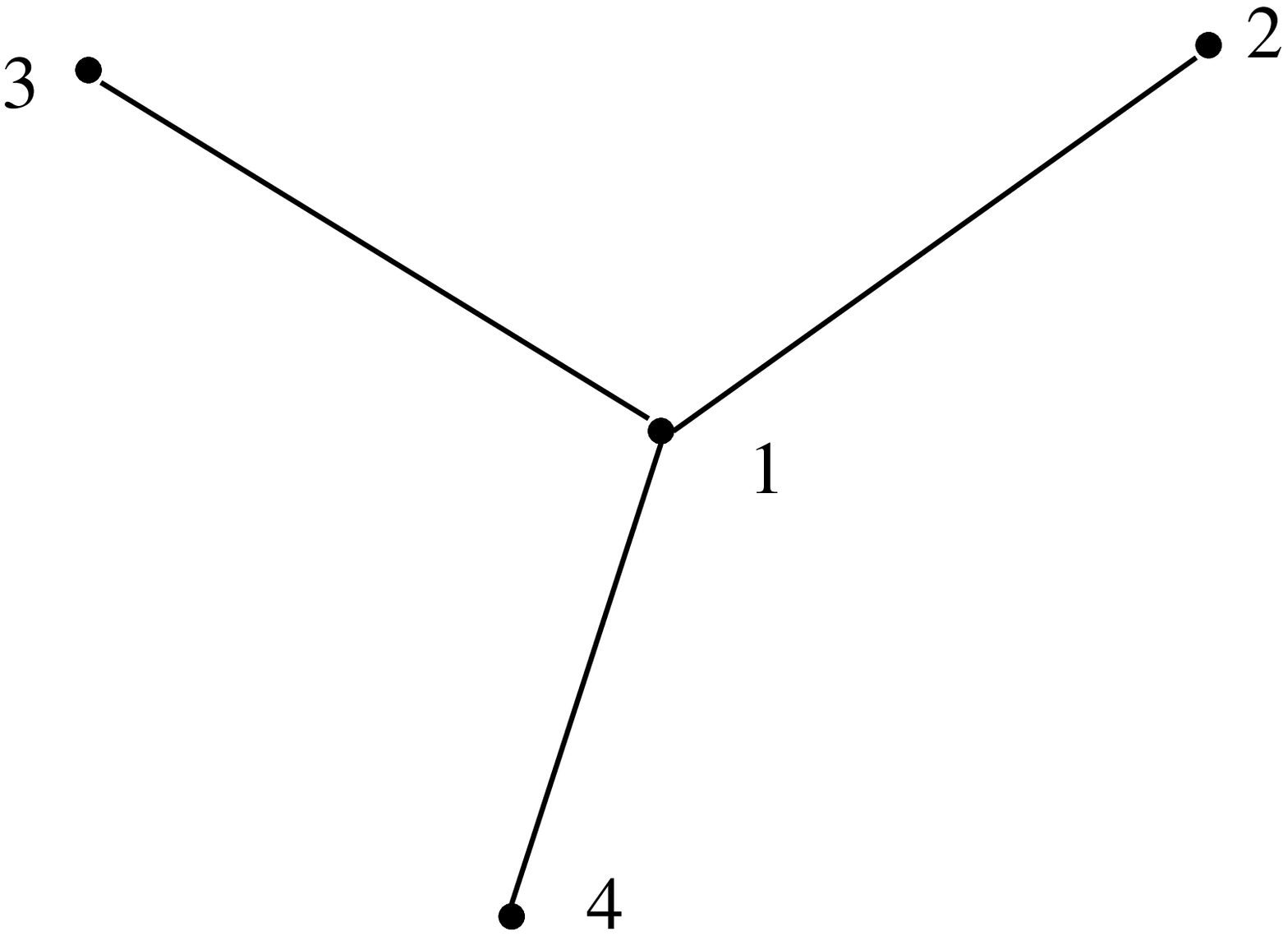}\hfil
\caption{Parts of the surface of the polytope $\D^*$ }
\label{fig:edge}
\end{figure}

Let us assume now that the fan $\S$ is given by a triangulation of the 
faces of $\D^*$ (we will see later that it is irrelevant which 
triangulation we choose) into elementary simplices (lattice simplices
containing no lattice points except for their vertices).
As any elementary triangle is also regular (of volume one in lattice units), 
the fan $\S$ consists
of cones whose integer generators $v_i$ generate the full lattice, implying
that the ambient space $X(\S)$ is smooth.
We will  use the same symbol $v_i$ both for lattice vectors and
for points determined by these vectors.
Triple intersections of three different toric divisors in $X(\S)$ are 
one if these divisors form a cone and zero otherwise.
For calculating intersections in the $K3$ surface, we have to 
evaluate expressions of the type $D_1\cdot D_2\cdot K3$ in the ambient space.
If all of the divisors of the $K3$ are intersections of divisors
of the ambient space with the $K3$ surface, this is all we need.
The $K3$ surface, as a divisor in the ambient space, is linearly equivalent 
to $\sum D_i$, so this task can be
reduced to calculations of the type $D_1\cdot D_2\cdot D_3$ in $X(\S)$.
In particular, it is clear that $D_1\cdot D_2\cdot K3$ can be nonzero
only if $v_1$ and $v_2$ belong to the same cone in $\S$ (the same triangle
on the surface of $\D^*$).
Let us first assume that $v_1$ and $v_2$ are distinct.
Then we have the situation shown in the first picture of figure 
\ref{fig:edge}, 
which depicts the part of the surface of $\D^*$ that is relevant for the 
calculation of $D_1\cdot D_2\cdot K3$.
It is obvious that no divisors except $D_1,\ldots, D_4$ are involved
in the calculation, so 
\beq D_1\cdot D_2\cdot K3=D_1\cdot D_2\cdot (D_1+D_2+D_3+D_4).\eeq
Let $m_{123}$ be the element of the $M$ lattice dual to the plane
that is affinely spanned by $v_1,v_2,v_3$.
Then we know that $\sum \<m_{123},v_i\>D_i\sim 0$.
With $\<m_{123},v_i\>=-1$ for $i=1,2,3$ we get
$D_1+D_2+D_3\sim \<m_{123},v_4\>D_4+\cdots$, where we have omitted
divisors that don't intersect $D_1$ or $D_2$.
This gives 
\beq D_1\cdot D_2\cdot K3=\<m_{123},v_4\>+1.\eeq
This formula has a nice interpretation: 
We notice that $\<m_{123},x\>+1=0$ is just the equation for the plane
carrying $v_1,v_2,v_3$.
Thus $\<m_{123},x\>+1=n$ is the equation for a plane at integer
distance $|n|$ from the one spanned by $v_1,v_2,v_3$.
Given that the triangle $v_1v_2v_3$ is regular, we see that
$\<m_{123},v_4\>+1$ is just the volume of the tetrahedron $v_1v_2v_3v_4$.
In particular the intersection of $D_1$ and $D_2$ in the $K3$
surface is zero whenever $v_1,v_2,v_3,v_4$ are in a plane, i.e.
when $v_1v_2$ is not part of an edge of $\D^*$.
Conversely, if $v_1$ and $v_2$ are neighbors along an edge $\th_{12}^*$, then
$m_{123}$ and $m_{124}$ are vertices of $\D$ that define the dual edge 
$\th_{12}$ of $\D$.
This edge $\th_{12}$ may or may not contain lattice points in its interior.
As $v_1,v_2,v_3$ generate $N$, a vector $m\in M_\IR$ is integer if and
only if $\<m,v_i\>$ is integer for $i=1,2,3$.
Given this, it is easily checked that the integer points of $\th_{12}$
are precisely the points 
$(km_{123}+(n-k)m_{124})/n$, where $n$ is the volume of the tetrahedron 
$v_1v_2v_3v_4$ and $k=0,1,\ldots,n$.
This means that $n$ is the length (in lattice units) $l_{12}$ 
of the edge $\th_{12}$ of $\D$ dual to the edge ${v_1v_2}$, i.e. that 
\beq D_1\cdot D_2\cdot K3=l_{12}=l'(\th_{12})+1, \eeq
where $l'(\th_{12})$ is the number of interior points of $\th_{12}$.

Let us now turn our attention to self-intersections $D_1\cdot D_1\cdot K3$.
We start with an observation concerning self-intersections of curves
in $K3$ surfaces that is not restricted to the toric case.
By the adjunction formula, the first Chern classes of the tangent and
normal bundles of a curve in a $K3$ surface must add up to the first 
Chern class of the $K3$, i.e. to zero.
For a curve embedded algebraically in a surface,
the self-intersection is given by the first Chern class
of its normal bundle, and the first Chern class of the tangent bundle
of a curve is just its Euler characteristic $\chi$.
Thus the self-intersection of an algebraic curve in a $K3$ surface is $-\chi$.
In particular, rational curves have self-intersections of $-2$ and elliptic 
curves have vanishing self-intersections.

Returning to the toric case,
let $m\in M$ be dual to a plane bounding $\D^*$ and containing $v_1$
(the lattice vector corresponding to $D_1$).
Then $D_1\sim \sum_{i>1}\<m,v_i\>D_i +\cdots$, so 
\beq 
D_1\cdot D_1\cdot K3 = \sum_{i>1}\<m,v_i\>D_1\cdot D_i\cdot K3  +\cdots.
\eeq
We can now use the knowledge we just gained about intersections of different
divisors in the hypersurface.
In particular, if $v_1$ is in the interior of a face, we see that its
self-intersection is 0.
If $v_1$ is in the interior of an edge, and if its neighbors on the
edge are $v_2$ and $v_3$, then we know that $D_2$ and $D_3$ are the
only divisors that intersect $D_1$. 
Moreover, they must necessarily lie in the plane $\<m,x\>+1=0$, so
\beq 
D_1\cdot D_1\cdot K3 = -D_1\cdot D_2\cdot K3 -D_1\cdot D_3\cdot K3 =-2l,
\eeq
where $l$ is the length of the edge dual to the one carrying $v_1,v_2$ and 
$v_3$.
By the general discussion above, we conclude that $D_1$ must be reducible
whenever $l>1$.
We note that this happens only when both an edge and its dual have interior
points, i.e. when the correction term in eq. (\ref{pic}) is nonzero.

Similar arguments may be used for self-intersections of divisors corresponding
to vertices. 
If $v_1$ is a vertex from which 3 edges originate, we have a situation as
depicted in the second picture in figure \ref{fig:edge}. 
With arguments as before, 
\beq D_1\cdot D_1\cdot K3 = -D_1\cdot D_2\cdot K3 -D_1\cdot D_3\cdot K3 -
     D_1\cdot D_4\cdot K3+(\<m_{123},v_4\>+1)D_1\cdot D_4\cdot K3. \eeql{trip}
To see that this expression is invariant under
permutations of $v_1,v_2,v_3$, we may note that the volume of the tetrahedron
$v_1v_2v_3v_4$ can be described alternatively as 
$(\<m_{123},v_4\>+1)a_{123}$ or as $l_{14}a_{124}a_{134}$,
where $a_{134}$ is the area (in lattice units) of the triangle
$v_1v_3v_4$, or by any expression obtained from one of these by 
permuting the labels 2,3,4.
These identities also allow us to give a bound on $D_1\cdot D_1\cdot K3$:
We may use them to rewrite (\ref{trip}) as
\beq D_1\cdot D_1\cdot K3 = -l_{12}-l_{13}-l_{14}+a_{134}l_{13}l_{14}.  \eeq
Assuming, without loss of generality (otherwise permute among 2,3,4), that 
$l_{12}\le l_{13}\le l_{14}$, it is easily checked that 
\beq 
-l_{12}-l_{13}-l_{14}+a_{134}l_{13}l_{14}\ge l_{12}(l_{12}a_{134}-3)\ge -2,
\eeq
the last inequality being true for any positive integer values of $l_{12}$ 
and $a_{134}$.

\begin{figure}[htb]
\epsfxsize=1.8in
\hfil\epsfbox{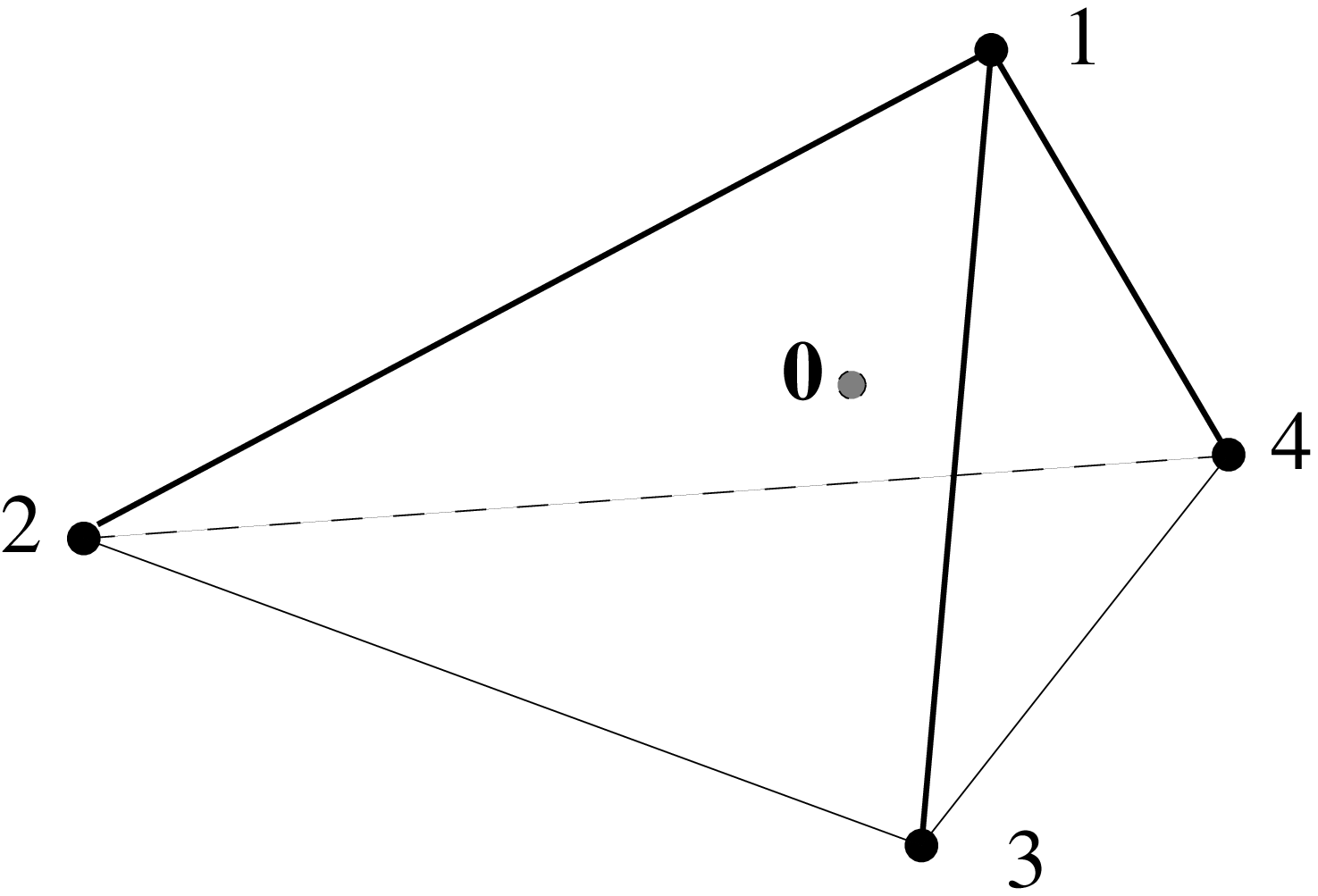}\epsfxsize=1.8in
\hfil\epsfbox{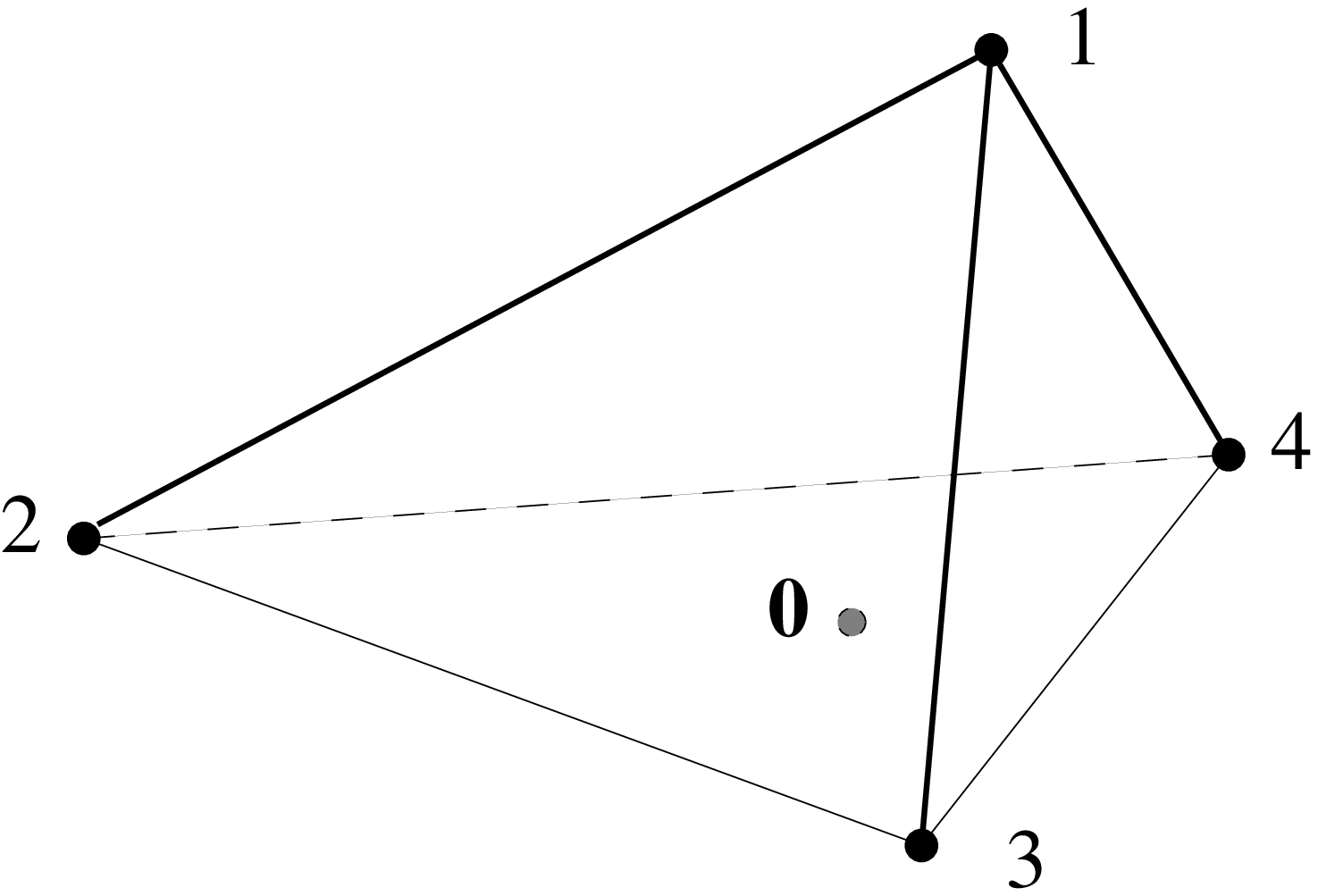}\epsfxsize=1.8in
\hfil\epsfbox{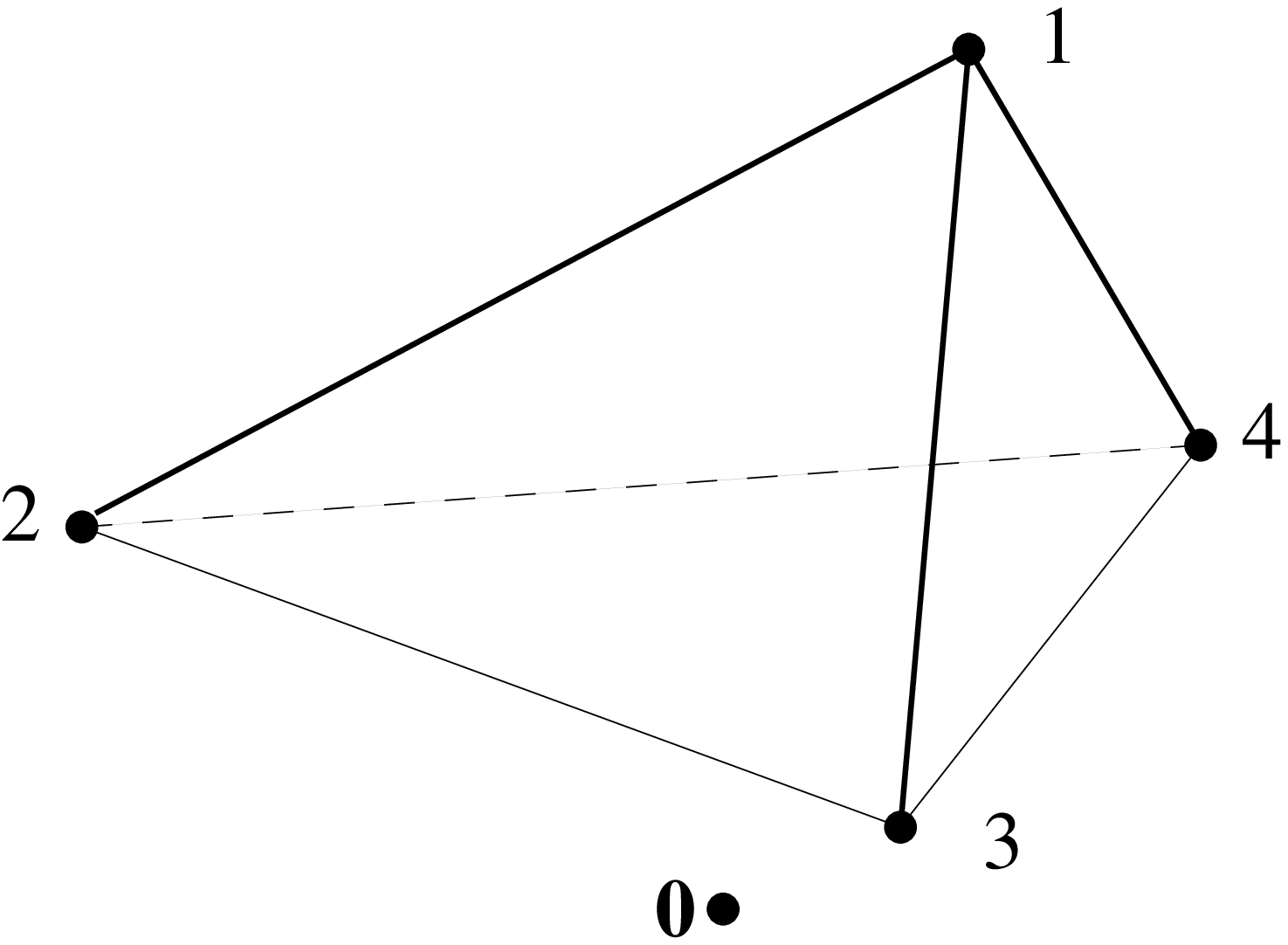}\hfil
\caption{Self-intersection for the vertex $v_1$}
\label{si}
\end{figure}

For a slightly different way of obtaining the self-intersection of $D_1$ 
consider figure \ref{si} (the lines not originating in $v_1$
serve for better visualization and need not correspond to parts of the
triangulation).
Now we use the equation $\<m_{234},v_i\>=a$ of the plane through
the $v_i$ with $i=2,3,4$.
Here the r.h.s. need not equal $-1$ because this plane need not bound $\D^*$.
With $\<m_{234},v_1\>=b$ we get
\beq D_1\cdot D_1\cdot K3 = -{a\0 b} (D_1\cdot D_2\cdot K3 +
     D_1\cdot D_3\cdot K3 +D_1\cdot D_4\cdot K3). \eeq
Corresponding to the three different situations depicted in figure
\ref{si}, we have the following possibilities:
$D_1\cdot D_1\cdot K3$ is positive if $\ipo$ and
$v_1$ are on the same side of the plane spanned by $v_1,v_2,v_3$.
An example for this case is $X(\S)=\IP^3$.
Here $v_1,v_2,v_3,v_4$ and $\ipo$ are the only lattice points in $\D^*$.
All $D_i$ are linearly equivalent, so their mutual and self-intersections are 
equal and are easily found to be $+4$.
$D_1\cdot D_1\cdot K3$ vanishes if $\ipo$ lies in the plane spanned by 
$v_1,v_2,v_3$ and is negative if $\ipo$ and $v_1$ are on different sides
of this plane.
In the latter case our previous arguments lead to the conclusion that we
are dealing with a single rational curve of self-intersection $-2$.
The same argument would apply with more than three neighbors of
$v_1$ as long as they are in a plane.
Other self-intersections of divisors corresponding to vertices with more  
edges originating from them can be calculated by similar means.

Let us summarize the results of this section: 
A divisor $D$ of the ambient space corresponding to a point $v$ of
the polyhedron $\D^*$ does not intersect the $K3$ hypersurface if $v$ 
lies in the interior of a facet of $\D^*$.
Mutual intersections of divisors $D_1$ and $D_2$ are nonzero if and only
if the corresponding points $v_1$ and $v_2$ are neighbors along
an edge $\theta^*$ of $\D^*$.
In that case the intersection number is equal to the length $l$ 
(in lattice units) of the dual edge $\theta$ in $\D$.
Self-intersections of divisors corresponding to points interior to
edges are equal to $-2l$, where $l$ is again the length of the dual
edge.
If $l>1$, such a divisor must be reducible.
Self-intersections of divisors corresponding to vertices are positive 
in the first case of figure \ref{si}.
They are 0 in the second case; here we are dealing with elliptic curves.
Finally, they are equal to $-2$ in the third case; this is the case of a 
rational curve.
\newpage

\section{Elliptic $K3$ Surfaces}

\subsection{Toric description of fibrations}

Now we turn our attention to $K3$ surfaces that are elliptic fibrations.
As shown in \cite{k3,fft}, this means that we have a distinguished direction
in the $M$ lattice, given by a primitive vector $m\fib$, which
determines a distinguished linear hyperplane 
\beq 
N\fib=\{y\in N: \<y,m\fib\>=0\}
\eeq
in the $N$ lattice such that 
\beq 
\D^*\fib:=\D^*\cap N\fib  
\eeq
is reflexive.
$N\fib$ divides $\D^*$ into an `upper' and a `lower' half to which we
will refer as `top' and `bottom', respectively.
The base space of the fibration is a $\IP_1$ with homogeneous
coordinates $(z_{\rm upper}:z_{\rm lower})$, where
\beq 
z_{\rm upper}~~=\prod_{i: \<v_i,m\fib\>>0}z_i^{\<v_i,m\fib\>},~~~~~~~~~
z_{\rm lower}~~=\prod_{i: \<v_i,m\fib\><0}z_i^{-\<v_i,m\fib\>}.   
\eeq
So the linear equivalence class of the fiber is given by
\beq 
D\fib~~ \sim \sum_{i: \<v_i,m\fib\>>0}{\<v_i,m\fib\>}D_i ~~~\sim~~~
      -\sum_{i: \<v_i,m\fib\><0}{\<v_i,m\fib\>}D_i .     
\eeql{Dfib}
Given this, it is easy to check that the self-intersection of
the generic fiber as well as the intersections of the generic fiber 
with any component of one of the exceptional fibers are zero.
What about the self-intersections of the exceptional fibers?
Self-intersections of divisors corresponding to interior points
of edges are of course always negative.
In all other cases we are dealing with vertices.
If there is only one upper point, this point simply corresponds to
the fiber without being exceptional; this is in agreement with our above 
analysis (self-intersection of a divisor corresponding to a point
whose neighbors form a plane that contains $\ipo$).
Let us now consider a vertex that is not the only upper point.
At least when there are only three neighbors,
they will form a plane that lies between the vertex we are considering
and $\ipo$, implying a self-intersection of $-2$.

\subsection{Weierstrass fibers}

Specializing further, we now assume that $\D^*\fib$ is a triangle
spanned by three vectors $v_x$, $v_y$ and $v_z=-2v_x-3v_y$ 
that generate $N\fib$.
Our toric variety will then be described by coordinates
\beq 
(x,y,z,z_1,\ldots)\sim (\l^2x,\l^3y,\l z,z_1,\ldots),  
\eeq
and the $K3$ surface is determined by a polynomial of degree 6 in $\l$.
With the usual redefinitions this polynomial can be chosen as
the Weierstrass polynomial in $x,y,z$ with coefficients that are functions
of $z_1,\ldots$.
We also assume that $v_z$ lies in an edge of $\D^*$, and that its neighbors 
along this edge are the points $v_a$ `above' and $v_b$ `below' $v_z$,
with $\<v_a,m\fib\>=1$, $\<v_b,m\fib\>=-1$.
This ensures, among other things, that the points interior to
the edges of the triangle $v_xv_yv_z$ are also interior to faces
of $\D^*$, so we can disregard them in the following analysis.
If there is no correction term, the Picard lattice is generated by the
divisors corresponding to points of $\D^*$ that do not lie on a facet 
modulo three relations of linear equivalence.
We may choose as the basis for the Picard lattice the
following divisors:
\beq D\fib,~~~~D_{\rm section}={D_z},~~~~\{D_{z_i}\}\setminus\{D_a,D_b\}.
\eeq
It is easily checked that this is a good basis (for example, one
may use linear equivalence to eliminate $D_x,D_y,D_a$ and 
trade $D\fib$ for $D_b$).
The divisors in this basis can be grouped into three distinct sets with
intersections only among members of the same set:
The first one is the system $\{D\fib,D_{\rm section}\}$ with
\beq 
D\fib\cdot D\fib=0,~~~D\fib\cdot D_{\rm section}=1,~~~
     D_{\rm section}\cdot D_{\rm section}=-2.   
\eeq
In addition there are the sets of divisors corresponding to upper 
points except $v_a$ and those corresponding to lower points except $v_b$.
Within each of these sets, different divisors intersect if and only
if they correspond to points that are connected by edges of $\D^*$.

So far we have talked only about smooth $K3$ surfaces. 
Let us now consider the scenario that we shrink to zero size all of
the divisors except for $D\fib$ and $D_{\rm section}$.
Thus we obtain a surface with two point singularities.
Such singularities have an $ADE$ classification which is reflected
in the intersection pattern of the divisors of its blow-up.
But this blow-up is just our original $K3$, and its intersection pattern
is given by the structure of edges.
We conclude that {\it the Dynkin diagrams of the gauge groups that appear
when the exceptional fibers are blown down to points are nothing but
the `edge diagrams' of the upper and lower parts of $\D^*$
without $v_a,v_b$, respectively}.

\begin{figure}[htb]
\epsfxsize=4in
\hfil\epsfbox{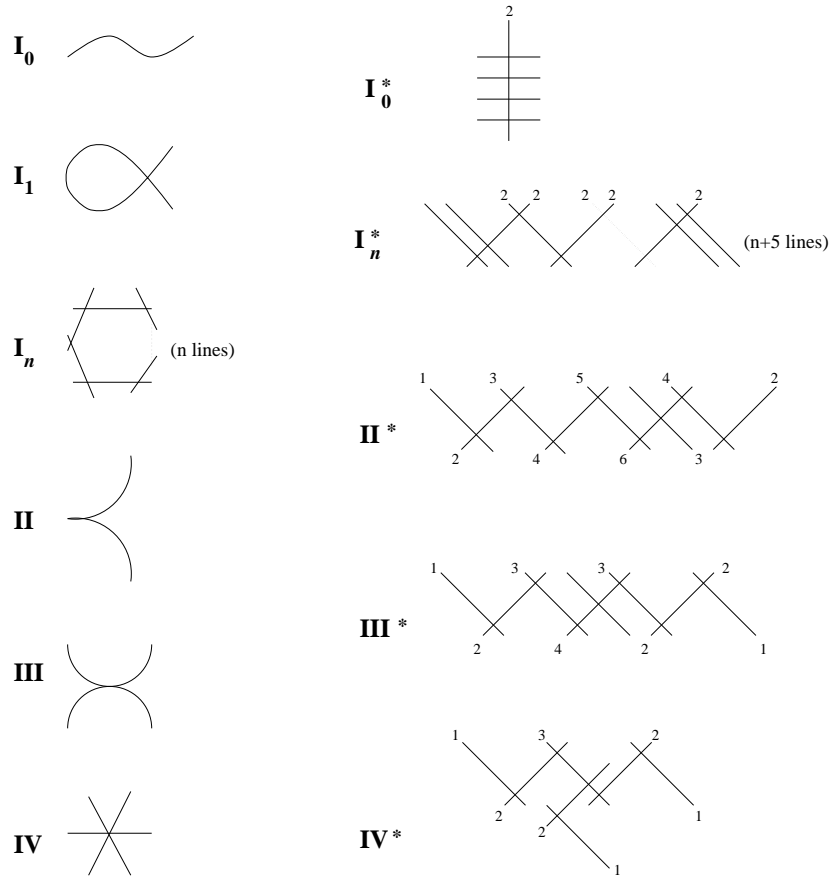}\hfil
\caption{Classification of elliptic fibers}
\label{classification}
\end{figure}

There is a slightly different way to find the singularity type given more
than one point projecting onto a given one-dimensional cone in 
$\S_{\rm base}$. It uses the Kodaira classification of degenerations of
elliptic fibrations \cite{Kod}. The possible types of degenerate fibers
are shown in figure~\ref{classification}  
and the gauge groups corresponding to those
fibers are presented in table~\ref{types}.

\begin{table}[htb]
\hfil
\begin{tabular}{|c|cccccccccc|}\hline
Fiber&I$_0$&I$_N$&II&III&IV&I$_0^{\ast}$&I$_{N-6}^{\ast}$&IV$^{\ast}$
&III$^{\ast}$&II$^{\ast}$\\ \hline
Singularity type&---&$A_{N-1}$&---&$A_1$&$A_2$&$D_4$&$D_{N-2}$&$E_6$&$E_7$
&$E_8$ \\ \hline
\end{tabular}
\hfil
\caption{The correspondence between 
Kodaira and $ADE$ classifications}
\label{types}
\end{table}

Each line on the figure (except for the $I_0$ case which is a generic smooth
elliptic fiber) represents a rational curve. In order for the 
exceptional fiber to be homologous to the generic fiber each rational curve
should be taken with multiplicity indicated by the number next to it.
Now notice that in our toric picture the class of the generic fiber is given
by eq. (\ref{Dfib}).
Interpreting the numbers $\pm\<v_i,m\fib\>$ as the multiplicities of rational
curves and comparing them to those shown in figure~\ref{classification} 
we can read off the type of the exceptional fiber occurring over the point 
$z_{\rm upper}=0$ ($z_{\rm lower}=0$) in the base. 
In other words, {\it the multiplicities shown in 
figure~\ref{classification} are nothing else than the heights of the 
corresponding points in the `top' in our toric picture}.

Let us briefly make the connection between the present description of 
singularities (via blowing down divisors) and the description that
is more common in the physics literature, namely by choosing 
particular equations. 
Consider the case where $\D^*$ has as vertices only $v_x$, $v_y$,
$v_a$ and $v_b$, i.e. where both `top' and `bottom' are trivial.
The most general equation (up to linear redefinitions of $x,y$) that 
would lead to a $K3$ hypersurface is then given by
\beq y^2=x^3+p_8(z_a,z_b)xz^4+p_{12}(z_a,z_b)z^6,  \eeql{triv}
where $p_8(z_a,z_b)$ and $p_{12}(z_a,z_b)$ can be arbitrary polynomials
of degrees 8 and {12}, respectively.
Upon restriction to 
\beq p_8=\a z_a^4z_b^4~~~~\hbox{ and } ~~~
     p_{12}=z_a^5z_b^7+\b z_a^6z_b^6+z_a^7z_b^5, \eeql{e8e8}
we get a homogeneous version of the equation considered in \cite{MVII}
for the description of an $E_8\times E_8$ symmetry. 
This can be explained in the following way:
When passing from the general form of eq. (\ref{triv}) to that determined 
by (\ref{e8e8}), we restrict $\D$ to a smaller polyhedron $\D'$.
The singularity coming from considering the non-dual pair $(\D^*,\D')$
may be resolved by blow-ups that change $\D^*$ to $(\D')^*$, so we
can read off the singularity type we get when we restrict eq. (\ref{triv}) 
to (\ref{e8e8}) by analyzing the structure of $(\D')^*$.
Indeed, as it should be, both the top and the bottom of $(\D')^*$
correspond to the `$E_8$ top' that we will shortly introduce.

\subsection{Examples of $ADE$ groups}

The `tops' for many $ADE$ series groups were constructed in \cite{CF}. 
There they were identified on the basis of the Hodge numbers of the 
resulting \CY\ threefolds and of Dynkin diagrams formed by the points in the
`tops'. Using the methods described in the present article we can calculate
intersection numbers for the divisors in the corresponding elliptic $K3$
and indeed confirm the statements made in \cite{CF}. 

\begin{figure}[htb]
\epsfxsize=2in
\hfil\epsfbox{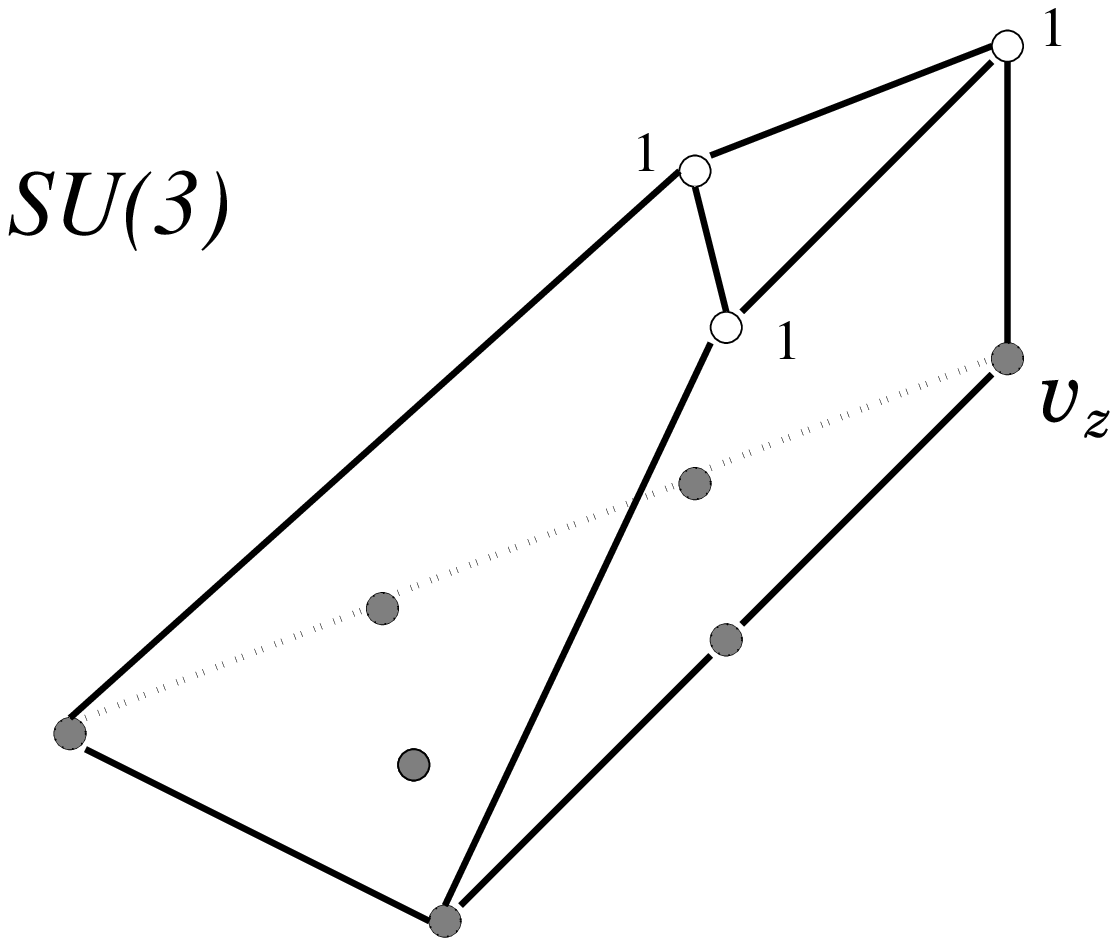}\hfil
\epsfxsize=2in
\hfil\epsfbox{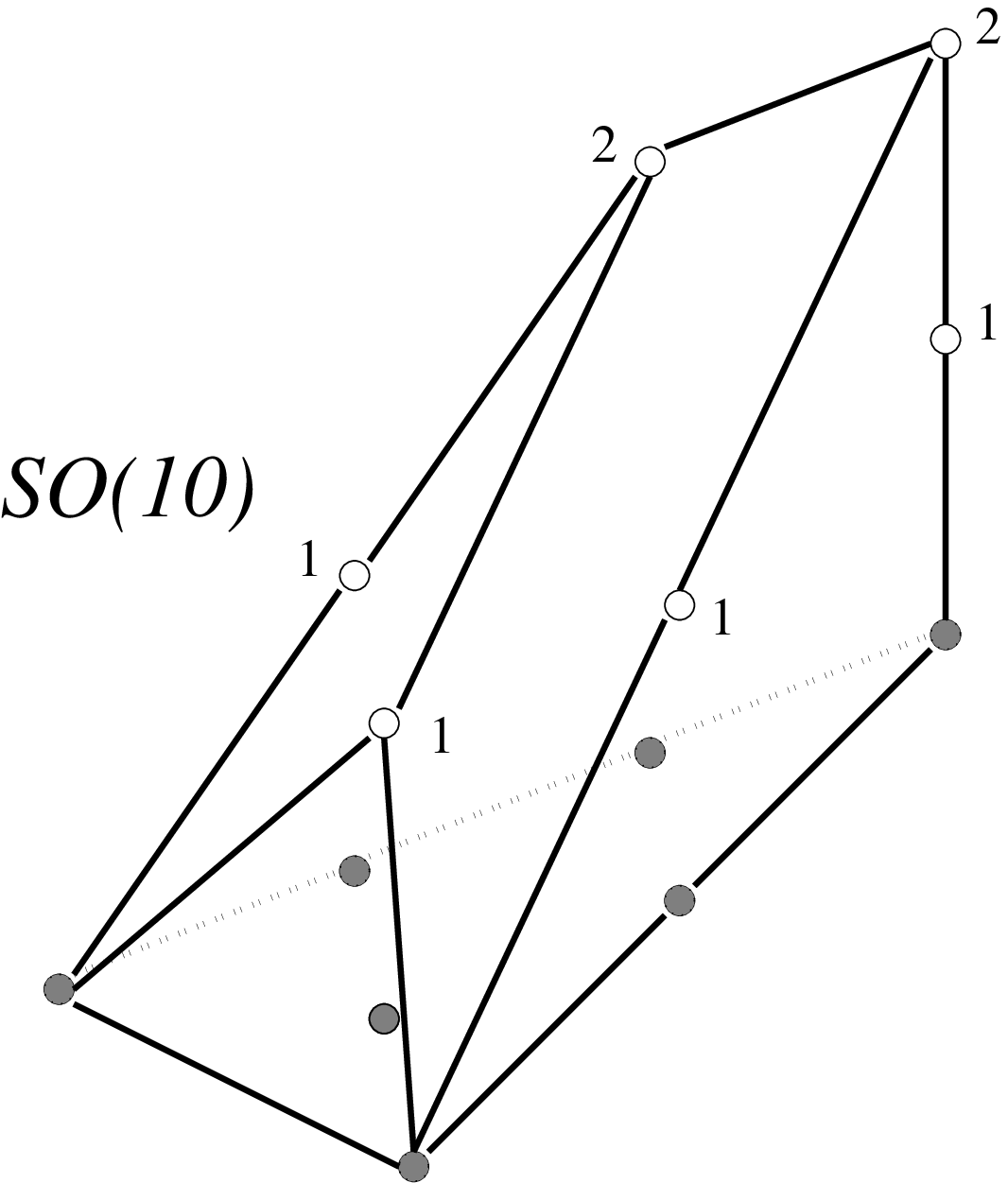}\hfil
\caption{The `tops' for $SU(3)$ and $SO(10)$.  The point $v_z$
is shown in the $SU(3)$ picture and is the same in the following examples. }
\label{su3}
\end{figure}

The `top' corresponding to the $SU(3)$ gauge group is shown in 
figure~\ref{su3}. 
The extended Dynkin diagram of $SU(3)$ is formed by the three white points 
one level above the plane of the generic fiber (grey points).
The intersection numbers between divisors in the $K3$ whose reflexive 
polyhedron can be formed by adding one point just below $v_z$ in the 
triangle of the generic fiber turn out to be 1 for each pair connected by an
edge. The self-intersections likewise are easily shown to be $-2$ for all
three points. This exactly corresponds to the occurrence of an $I_3$ fiber
in the elliptic $K3$ leading to an enhanced $SU(3)$ gauge symmetry in 
space-time if the rational curves in the degenerate fiber are blown down.
Note also that the three points describing the reducible fiber are all at 
height 1 above the triangle in precise agreement with the fact that the 
$I_3$ fiber is formed by three rational curves which should be taken with
multiplicity 1 each in order for the degenerate fiber to be homologous to the
generic one.

We illustrate the $D$ series case by the $SO(10)$ example shown in 
figure~\ref{su3}. There are 6 points in the `top' forming an extended
Dynkin diagram of $SO(10)$ (as in all the rest of our examples, we do not
show the points interior to codimension one faces in the `tops' since, as
was explained earlier, the corresponding divisors in the ambient space do not
intersect the hypersurface we are interested in). The divisors corresponding
to points joined by the edges intersect with intersection number 1 for all 
pairs. The self-intersections found from figure~\ref{su3} are all $-2$, 
precisely what we expect for an $I_1^{\ast}$ fiber. Note again that two points
in the middle of the extended Dynkin diagram at hand are at height 2 above
the triangle, all the rest of them being at height 1. These numbers follow 
exactly the multiplicities pattern shown for the $I_1^{\ast}$ fiber in 
figure~\ref{classification}.

\begin{figure}[htb]
\epsfxsize=2in\hfil\epsfbox{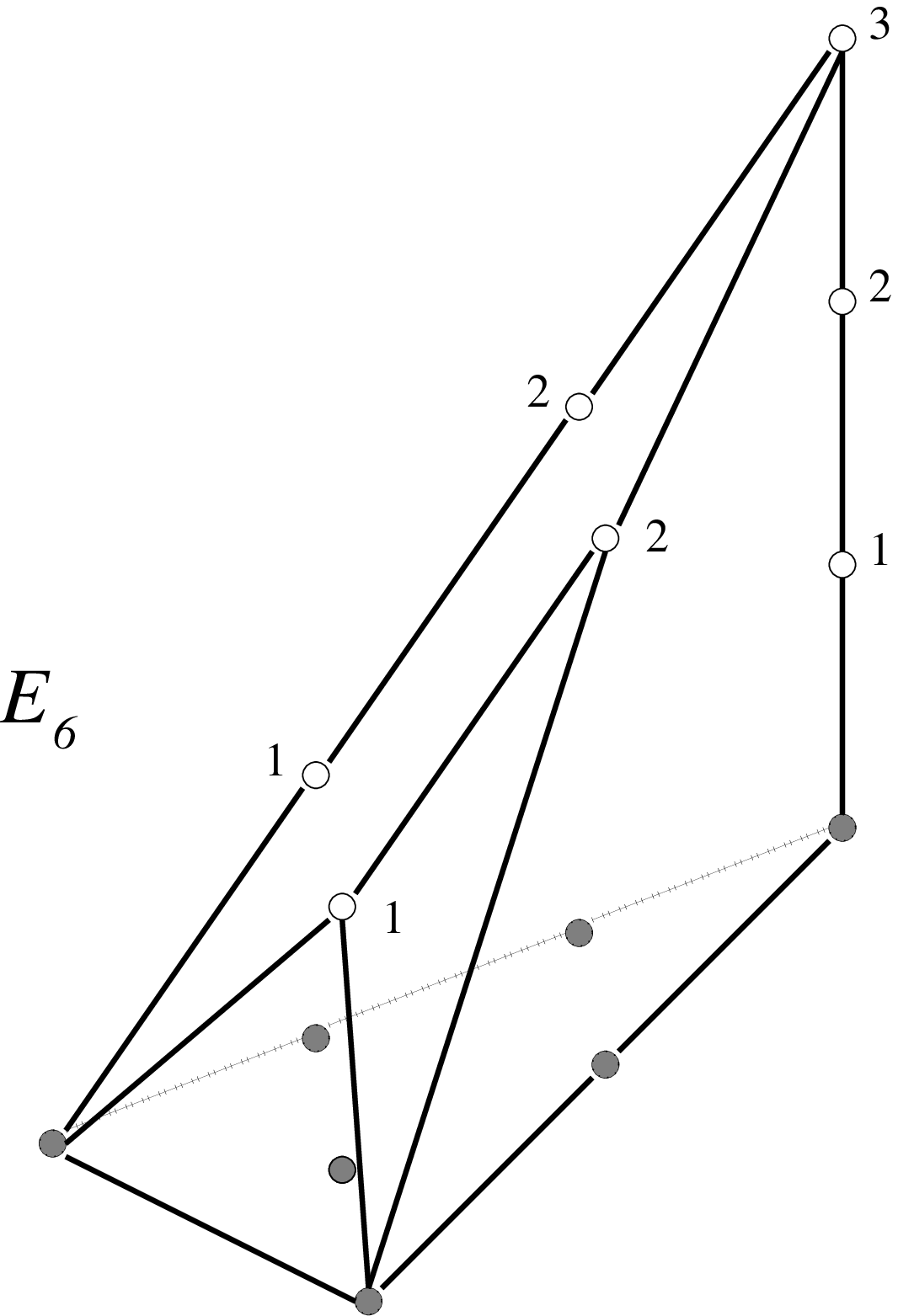}\hfil
\epsfxsize=2in\hfil\epsfbox{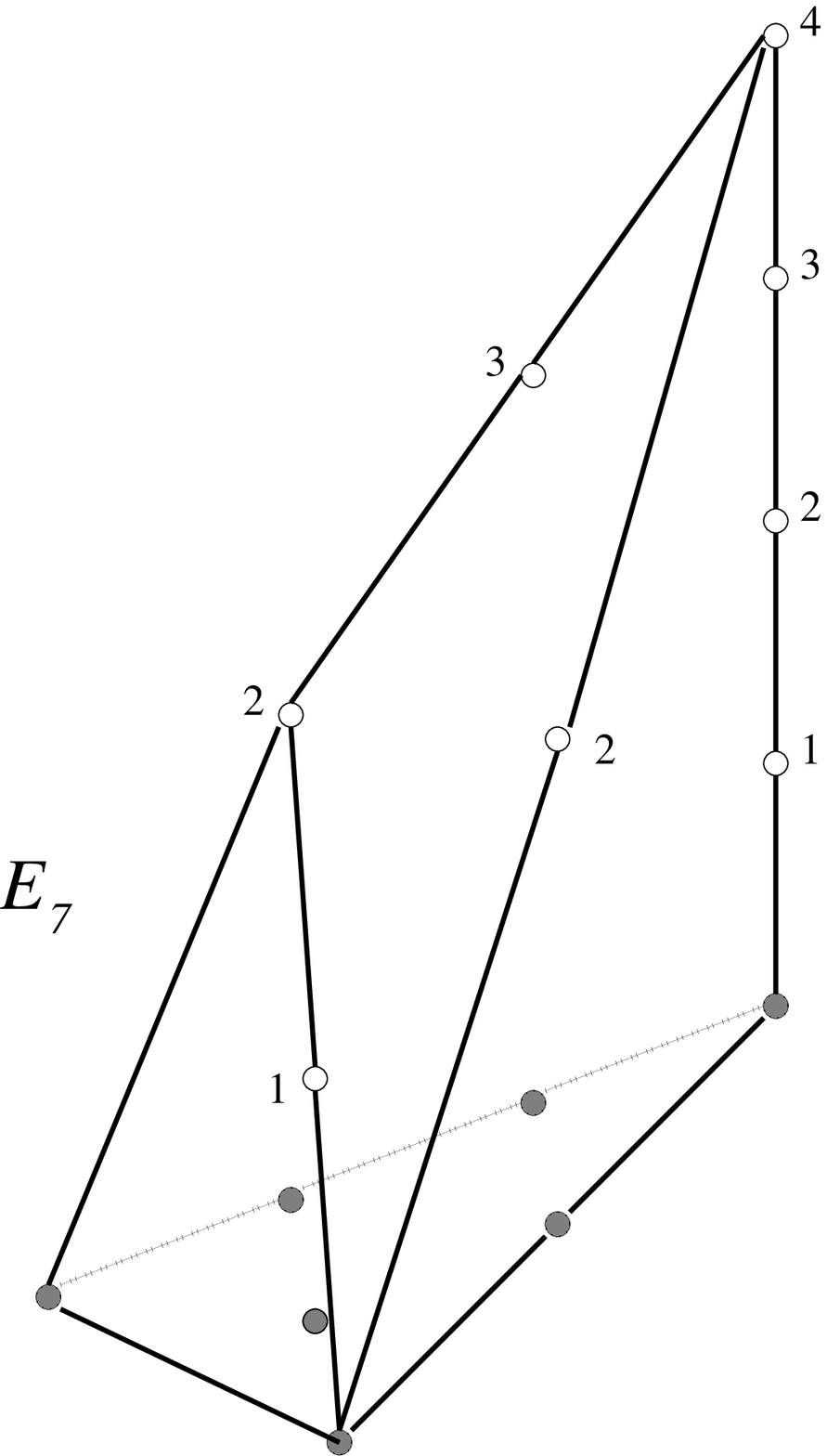}
\caption{The `tops' for $E_6$ and $E_7$}
\label{e67}
\end{figure}

\begin{figure}[htb]
\epsfxsize=2in
\hfil\epsfbox{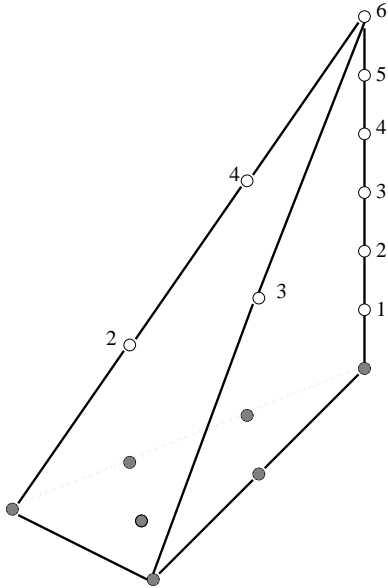}\hfil
\caption{The `top' for $E_8$ (the vertical scale is half of
that for the rest of `tops' figures)}
\label{e8}
\end{figure}

The `tops' corresponding to $E$ series groups are shown in figures~\ref{e67}
and~\ref{e8}. The divisors corresponding to points joined by an edge intersect
and all intersection numbers are indeed found to be 1, as well as all
self-intersection numbers turn out to be $-2$ revealing the intersection
patterns of rational curves constituting $IV^{\ast}$, $III^{\ast}$ and 
$II^{\ast}$ degenerations of elliptic fibers respectively. 
Extended Dynkin diagrams are hence
formed by those points, again confirming the observation made in \cite{CF}.
The heights of
the points above the plane of the triangle follow precisely the multiplicities
of rational curves shown in figure~\ref{classification}. It is also amusing
to calculate the ranks of Picard groups for the $K3$ surfaces described
by the reflexive polyhedra obtained by adding a trivial `bottom' (in 
terminology of \cite{CF}) to our `tops'. The application of (\ref{pic})
yields 8, 9 and 10 in the $E_6$, $E_7$ and $E_8$ cases, respectively, the 
correction (third) term in (\ref{pic}) being always zero.

\newpage

\section{Threefold Case}

\subsection{Toric description}

In this section, we generalize the construction presented in the previous
one to the case of elliptically fibered Calabi--Yau threefolds. Namely,
we present an explanation of the methods used in \cite{mon} to unravel
the singularity structure of elliptic Calabi--Yau threefolds given as
hypersurfaces in toric varieties. 

Again, as in the elliptic $K3$ case of the previous section, such a 
Calabi--Yau threefold being an 
elliptic fibration is equivalent to the existence of a two-dimensional 
plane $H$ in $N_\IR$ such that $H\cap \D^{\ast}=\D^{\ast}_{\rm fiber}$ 
is a reflexive 
polyhedron describing the generic fiber of the elliptic fibration. This 
observation was first made in \cite{CF}. As was shown in \cite{fft},
the base in this case can be seen by projecting the $N$ lattice along the 
linear space $H$ spanned by $\D^{\ast}_{\rm fiber}$. The projection map from 
$X(\S)$ (four-dimensional embedding variety of the Calabi-Yau threefold, in our
case) to the base was given as follows.

The set of one-dimensional cones in $\S\bas$ is the set of 
images of one-dimensional cones in $\S_{CY}$ that do not lie in $N\fib$. 
The image of a primitive generator $v_i$
of a cone in $\S_{CY}$  is the origin or
a positive integer multiple of a primitive
generator $\tilde v_j$ of a one-dimensional cone in $\S\bas$. 
Thus we can define a matrix $r^i_j$, most of whose elements are $0$, through
$\p v_i=r_i^j\tilde v_j$ with $r_i^j\in \IN$ if $\p v_i$ lies in
the one-dimensional cone defined by $\tilde v_j$ and $r_i^j=0$ otherwise.
Our base space is the multiply weighted space determined by
\beq 
(\tilde z_1,\ldots,\tilde z_\Nt)\sim (\l^{\tilde w^1_j}
\tilde z_1,\ldots,\l^{\tilde w^\Nt_j}\tilde z_\Nt),~~~~~~
j=1,\ldots,\tilde N-\tilde n  
\eeql{ber} 
where the $\tilde w^i_j$ are any integers such that 
$\sum_i\tilde w^i_j \tilde v_i=0$.
The projection map from $X(\S)$ (and also, as was demonstrated in
\cite{fft}, from the Calabi--Yau hypersurface) to the base is given by
\beq 
\tilde z_i=\prod_jz_j^{r_j^i}. 
\eeq

The fiber can degenerate. There are two possible mechanisms for that.
Since the fiber is a hypersurface in $X(\S_{\rm fiber})$, it can happen either 
when the embedding variety  $X(\S_{\rm fiber})$ itself degenerates or the 
equation of the hypersurface becomes singular. 
As all divisors are manifest in the blown-up toric picture, the second case can
only correspond to the cases $I_1$ or $II$ in the Kodaira classification,
which do not lead to enhanced gauge groups.
If a one-dimensional cone with primitive generator $\tilde{v}_i$ in
$\S_{\rm base}$ is the image of more than one one-dimensional cone in $\S$,
the fiber over the divisor $\tilde D_i$ determined by $\tilde{z}_i=0$ is 
reducible: different components 
of the fiber corresponding to equations $z_j=0$ project on $\tilde D_i$
whenever different cones $v_j$ project on one cone $\tilde v_i$ in 
$\S_{\rm base}$. 

If we take a small disk $D\subset\IC$ in the base that intersects the divisor 
$\tilde D_i$ transversely at a generic point 
we can consider the picture locally near the point of intersection $p$.
To determine the singularity type along the divisor $\tilde D_i$ we need to 
find the intersection numbers of the components of the fiber over $p$ in the 
two-dimensional manifold which is a fibration over $D$. To accomplish this 
task by our toric means we consider only the points of $\D^{\ast}$ which
project to the one-dimensional cone in $\S_{\rm base}$ corresponding to 
$\tilde D_i$. These points form a three-dimensional fan describing the local 
data we are after. Indeed, the divisor $\tilde D_i$ is given by 
$\tilde{z_i}=0$, where $\tilde{z_i}$ is the coordinate assigned to the 
one-dimensional cone corresponding to that divisor. Taking only that 
one-dimensional cone
means that the only coordinate we are left with is $\tilde{z_i}$. Varying it
we move in a direction in the base transverse to our divisor.
What we get is in fact one of the `tops' discussed in section 3, with the 
`heights' determined by the numbers $r_j^i$. The computation of the 
intersections presented there carries over to the threefold case.

\subsection{Monodromy and non-simply laced groups} 

\begin{figure}[htb]
\epsfxsize=4in
\hfil\epsfbox{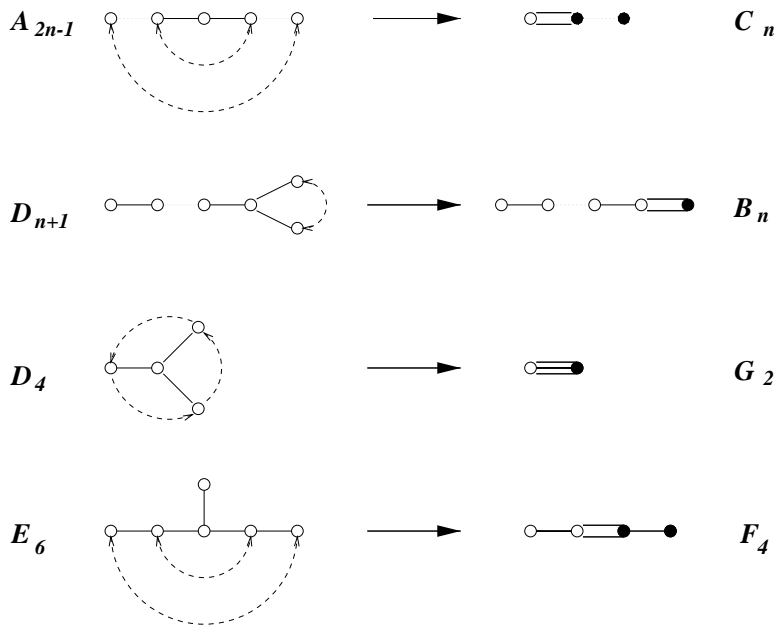}\hfil
\caption{Outer automorphisms of Lie algebras}
\label{outer}
\end{figure}

Up until now the enhanced gauge symmetry occurring in the uncompactified 
dimensions
coincided with the corresponding singularity type of the internal manifold.
This does not have to be the case though as was first shown in \cite{AG}. 
It is indeed so provided the rational curves within the exceptional fiber
are monodromy invariant as we move around the corresponding curve within the
base. In the cases when those curves are not monodromy invariant, the gauge 
group appearing in space-time is actually a subgroup invariant under
the outer automorphism obtained by translating the monodromy action on the 
rational curves into an action on the Dynkin diagrams of the simply-laced
gauge group corresponding to the singularity type. The possible outer 
automorphisms and their actions on the Dynkin diagrams are shown in 
figure~\ref{outer}. 

What happens from the toric point of view is the following. If there is 
a monodromy action on the rational curves in the fiber over a point $p$ 
then several divisors in the $K3$ which are exchanged by the monodromy 
action yield
only one divisor in the \CY\ threefold when transported over the whole
of $\tilde D_i$, the divisor in the base. 
Hence, they are represented by just one
point\footnote{We imply that all divisors in the \CY\ are represented by points
in $\D^{\ast}$, i.e. that the correction (third) term in (\ref{pic}) 
vanishes (cf. the remark at the end of section 2).} 
in the polar polyhedron $\D^{\ast}$ of the \CY\ threefold. In the complex 
surface which is a fibration over the disk $D$, that point in $\D^{\ast}$ 
(when regarded
as a point in the particular `top', the inverse image of the cone $\tilde v_i$ 
in $\S\bas$) corresponds to several divisors or, more precisely, to their sum.
Hence, the self-intersection of the divisor class represented by this 
particular point in the `top' should turn out to be $-2m$, where $m$ is the
number of rational curves identified under the monodromy.  One more point is 
worth mentioning. By inspection of 
figures~\ref{classification} and~\ref{outer}, we can convince ourselves that 
the rational curves exchanged by a monodromy always have equal multiplicity.
Hence the points representing the divisor classes of their sum should
have the height in the corresponding `tops' equal to that multiplicity.
We will see  that this is indeed the case in the examples below.

\subsection{Examples of non-simply laced groups}

\subsubsection{$SO(2n+1)$}

An $SO(2n+1)$ gauge group appears as a result of a monodromy action on 
$I_{n-3}^{\ast}$ fibers which exchanges two rational curves leading to only
one divisor in the \CY\ threefold when transported over the divisor in the 
base of the elliptic fibration. Thus these two rational curves are represented
by one point in the polyhedron $\D^{\ast}$. This point when regarded
as a point in the corresponding `top' represents the sum of the two rational 
curves exchanged by the monodromy. Thus the self-intersection of the divisor
class should be $-4$. The intersection number between this divisor class and 
the  divisor class  corresponding to the point in the
Dynkin diagram of $SO(2n+2)$ connected to the two points exchanged by the outer
automorphism  equals $1+1=2$. The remaining intersections involve simply
rational curves and should be equal to $-2$ for self-intersections and 1 for
mutual ones. As an illustration to this case, we depict the `top' corresponding
to the gauge group $SO(9)$ in figure~\ref{sp3}. The self-intersection number
calculated for the black dot in the figure is $-4$ and the intersection number
between the black dot and the white one connected to it by an edge is 2. The 
other self-intersections are $-2$ as expected.
The heights of the white points are exactly as in $SO(10)$ case. The height
of the black point is 1 since the multiplicities of the rational curves
in the $I_1^{\ast}$ fiber exchanged by the monodromy are 1.  
As always, the extended Dynkin
diagram of $SO(9)$ is visible in the `top'. The $K3$ polyhedron obtained by 
adding one point just below $v_z$ has Picard number 6, in agreement with
the presence of an $SO(10)$ lattice in the Picard lattice.

\subsubsection{$Sp(n)$}

\begin{figure}[htb]
\hfil\epsfxsize=2in
\hfil\epsfbox{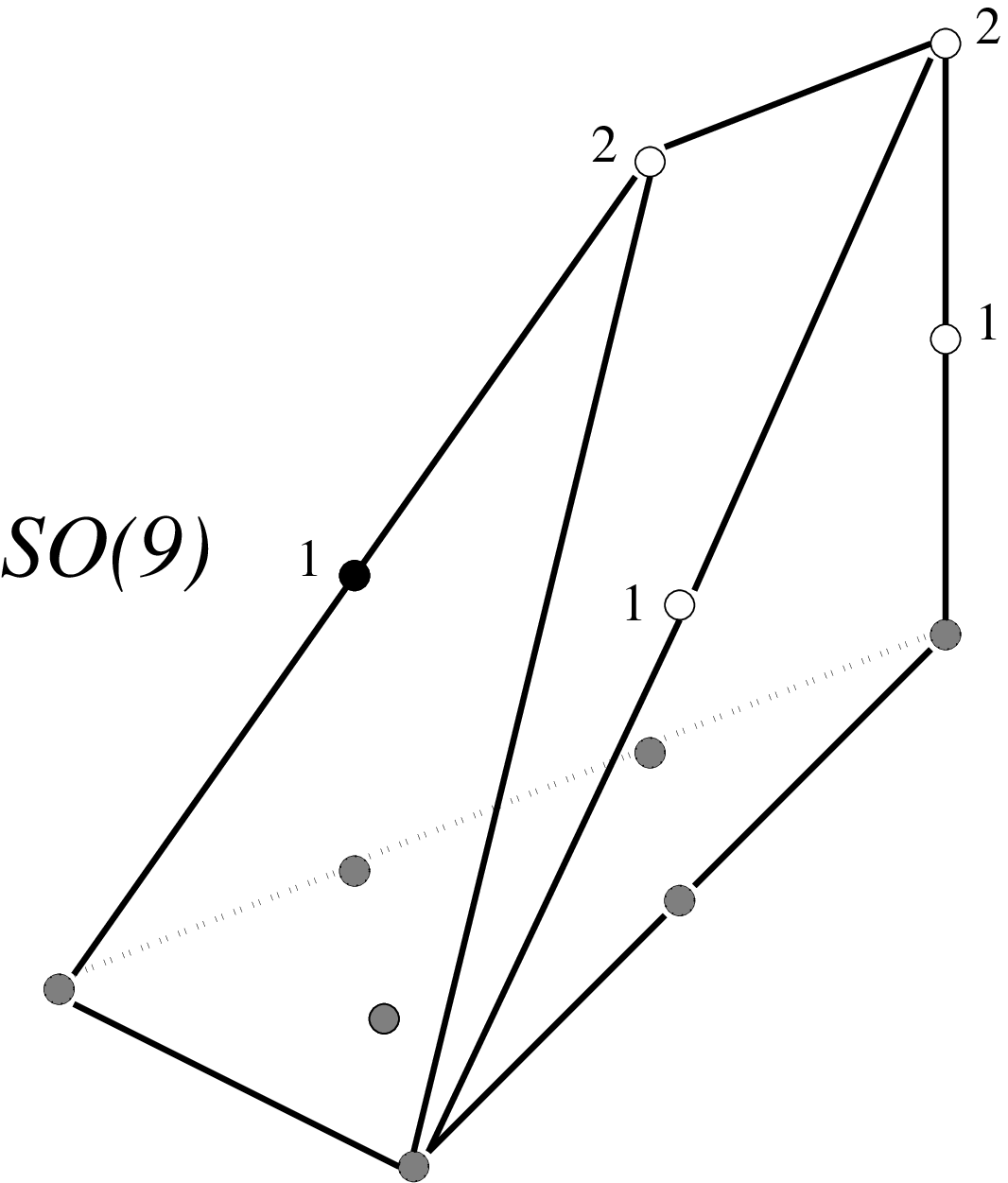}\hfil\epsfxsize=2in
\hfil\epsfbox{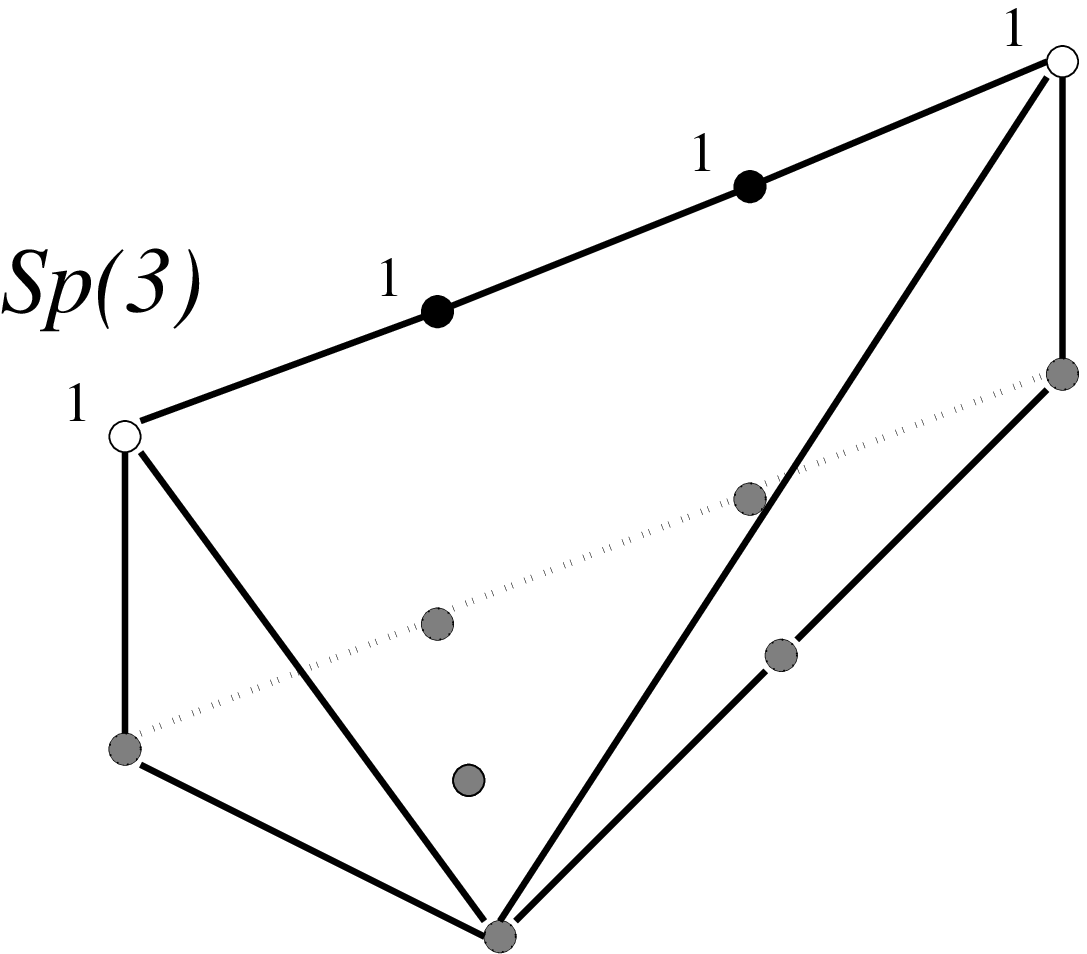}
\caption{The `tops' for the gauge groups $SO(9)$ and $Sp(3)$. }
\label{sp3}
\end{figure}

In order to obtain $Sp(n)$ symmetry in space-time we need a curve of 
$SU(2n)$ singularities in our \CY\ threefold, or, in other words, a divisor
in the base with $I_{2n}$ fibers over it subject to a monodromy action.
The pairs of curves exchanged by the monodromy will produce only one divisor
in the \CY\ each when transported over the divisor in the base. Hence, in 
our complex surface which is a fibration over the small disk $D$, the divisor
classes of the sums of the rational curves identified under the monodromy
will be represented by points in $\D^{\ast}$ of the \CY\ threefold. That is,
we expect $n+1$ relevant points in the corresponding `top', $n-1$ of them with
self-intersection $-4$ and all mutual intersections equal to 2. The 
self-intersections for the remaining two points are $-2$, each of them 
representing only one rational curve in the complex surface. We illustrate 
this case by an $Sp(3)$ `top' found in \cite{CF}. It is shown in 
figure~\ref{sp3}. As before, the points in the subpolyhedron of the generic
fiber are shown in grey color. All the relevant points are white and black, the
black ones corresponding to sums of the rational curves identified under the
monodromy. All the multiplicities in $I_{6}$ fibers are one. Hence the heights
of all points in the $Sp(3)$ `top' should be 1. We see that this is indeed 
the case. The extended Dynkin diagram of $Sp(3)$ is clearly visible
and appears in a natural fashion. 
As an additional check, note that adding one point below the point $v_z$ we 
obtain a reflexive polyhedron of a $K3$ surface whose Picard number equals 7,
in agreement with the presence 
of an $SU(6)$ gauge group (which gives $Sp(3)$ under the outer 
automorphism action).

\subsubsection{$F_4$}

\begin{figure}[htb]
\epsfxsize=2in
\hfil\epsfbox{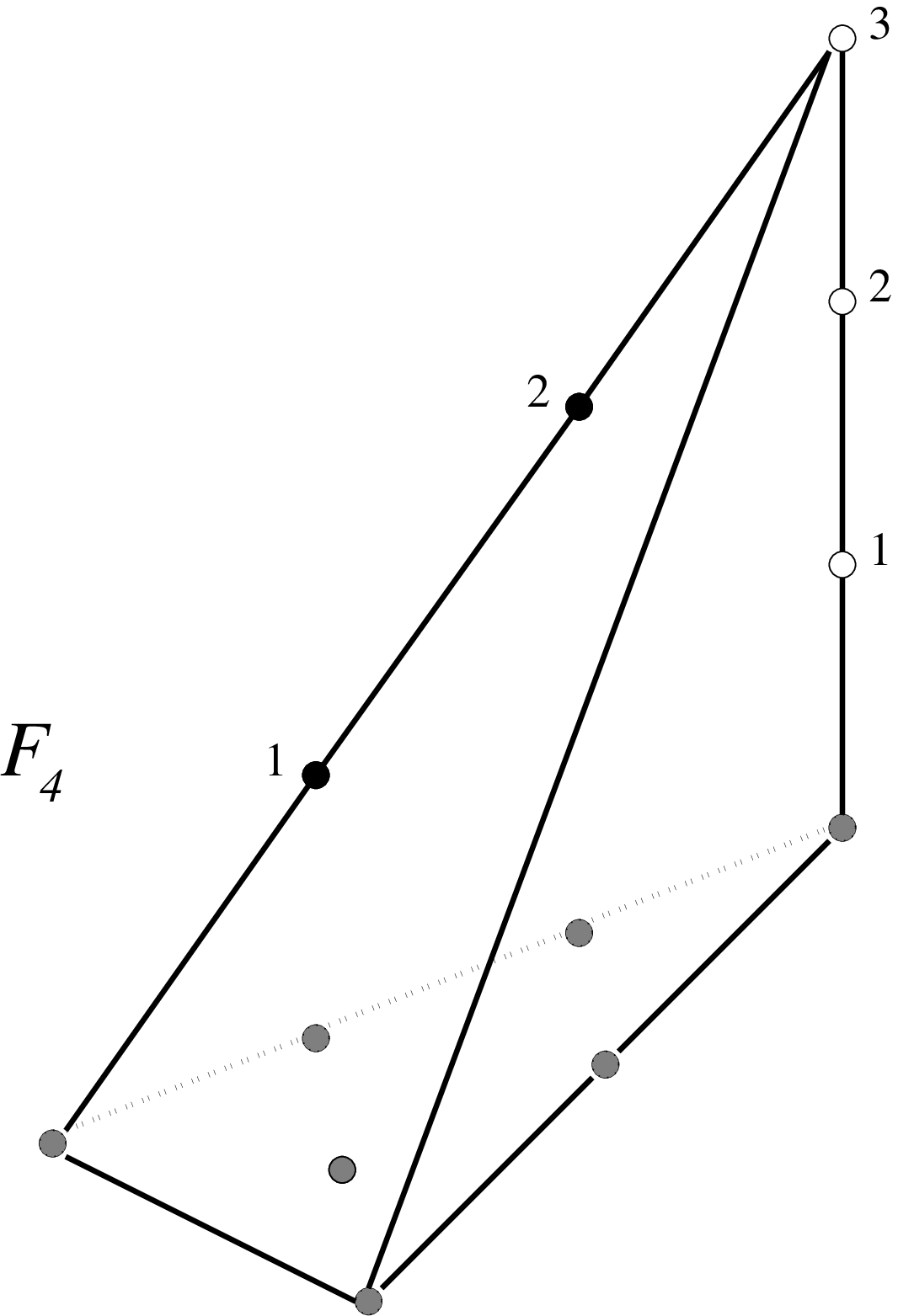}\hfil
\epsfxsize=2in
\hfil\epsfbox{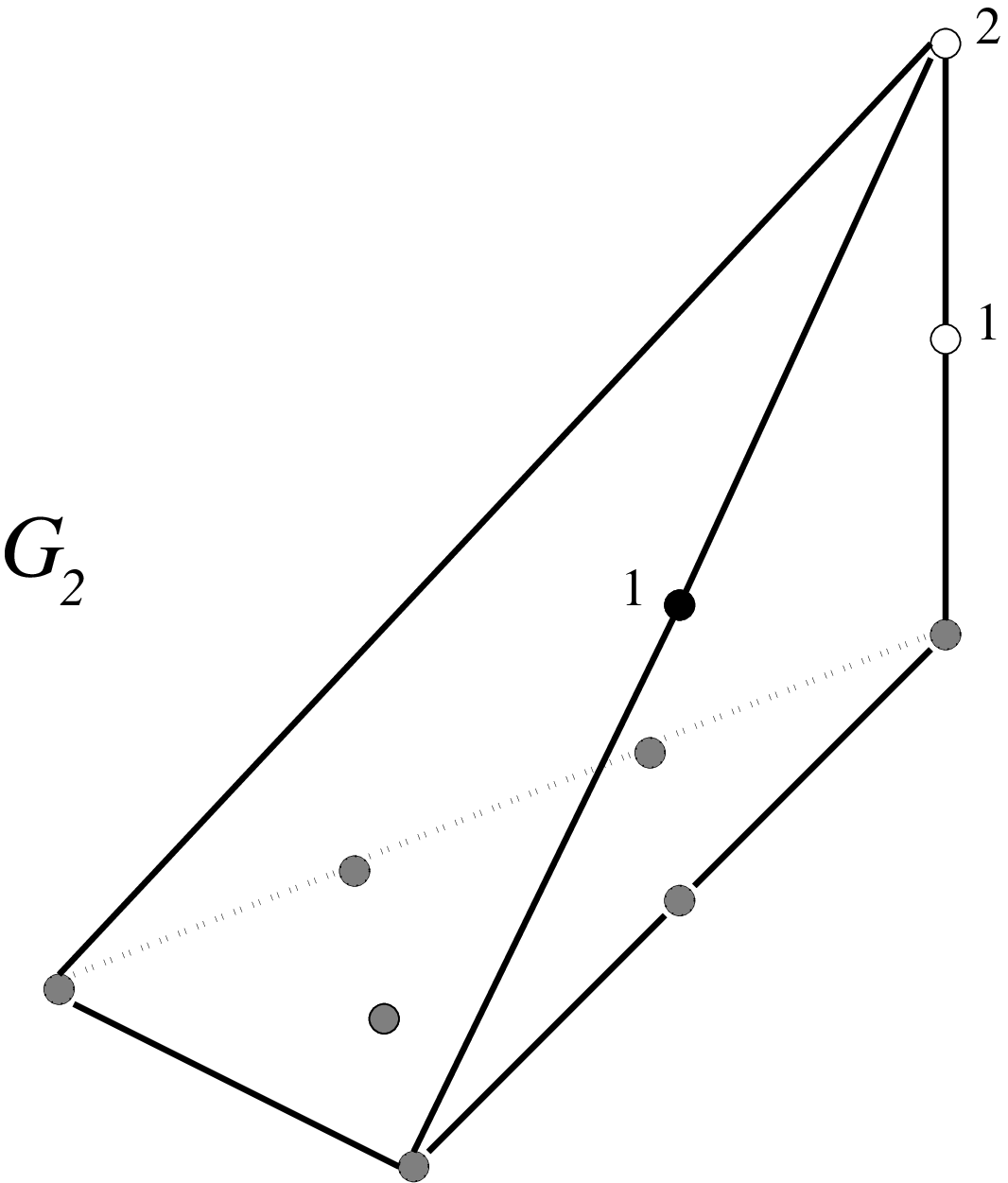}\hfil
\caption{The `tops' for $F_4$ and $G_2$ }
\label{g2}
\end{figure}

When the monodromy acts on a curve of $IV^{\ast}$ fibers $F_4$ gauge symmetry 
results. Two pairs of rational curves are exchanged under this monodromy
leading to two points in the \CY\ threefold which describe divisor classes
in the complex surface with self-intersections of $-4$. The self-intersection
of the remaining three rational curves in the $IV^{\ast}$ fiber we expect to be
simply $-2$. The mutual intersections involving divisor classes of sums of 
rational curves are easily seen to be 2, and 1 otherwise. This is exactly 
what we calculated in the $F_4$ `top' shown in figure~\ref{g2}. As before, 
black dots represent the sums of rational curves and white ones rational curves
forming the extended Dynkin diagram of $F_4$. Note also that the heights of 
the points shown in the figure are exactly as in the $E_6$ case except for the 
fact that two white points of height 2 and two white points of height 1
are replaced by one black point of the same height, respectively, following
the monodromy action on the rational curves. 
As in the previous case, we can 
construct a reflexive polyhedron of a $K3$ by adding one point to the `top'. 
The Picard number of this $K3$ turns out to be 8, as we could expect.

\subsubsection{$G_2$}

$G_2$ gauge symmetry is a result of a monodromy action on a curve of 
$I_0^{\ast}$ fibers. Under this monodromy three rational curves are exchanged,
so we expect the divisor class of their sum to be represented by one point
in the corresponding \CY\ threefold. The self-intersection of this divisor
class we expect to be $-6$. Correspondingly, the intersection number between 
this
divisor class and the divisor which is invariant under the monodromy should be
3. The intersection between the latter and another such (the one which
corresponds to the extending point in the extended Dynkin diagram of $G_2$)
should be 1. Of course the self-intersections of the latter two are $-2$ as 
they represent rational curves. The `top' corresponding to $G_2$ is shown in
figure~\ref{g2}. The black dot represents the sum of three rational curves.
The $I_0^{\ast}$ fiber has one rational curve with multiplicity 2. It is
invariant under the monodromy. Hence there is one white point of height 2
in our `top'. The three curves exchanged by the monodromy all have multiplicity
1. The black dot is at height 1 in agreement with that.
The extended Dynkin diagram of $G_2$ is again clearly visible. By adding one 
point below $v_z$ we obtain the reflexive polyhedron
of the $K3$. The calculation of the Picard number gives 6, in agreement
with the presence of $SO(8)$ in the Picard lattice.

\section{Discussion}

In this article we have given a systematic exposition of methods allowing
one to read off the singularity structure of \CY\ manifolds described as
hypersurfaces in toric varieties. We particularly focused on the cases leading
to enhanced gauge symmetries in F-theory (or type II) compactifications. 
These methods have already been used in the literature \cite{CF,mon} but an 
explanation was lacking. We have shown that $ADE$ orbifold singularities 
of a toric $K3$ surface (upon their resolution) exhibit themselves in the
form of extended Dynkin diagrams of corresponding simply laced Lie groups
appearing in the reflexive polyhedra, the fact noticed and used in \cite{CF}.
These are precisely the groups observed in the low energy spectrum of the
corresponding compactified theory. Moreover, we have shown that a nontrivial 
monodromy action in the case of a \CY\ threefold with curves of singularities
leading to the appearance of non-simply laced groups is also encoded in the 
corresponding polyhedra in a rather simple fashion. As also was noticed in
\cite{CF}, the extended Dynkin diagram of a simply laced group (the singularity
type along the curve) undergoes just the right outer automorphism to yield
the extended Dynkin diagram of the non-simply laced group which is observed in
the effective theory. 

These methods, though discussed in the case of elliptic \CY\ threefolds 
carry over without any major changes to the elliptic fourfold case (and, in
principle, to $n$-folds with $n>4$). Each ray in the fan of the base represents
a divisor and the points of the polyhedron projecting on a given ray encode 
the way the elliptic fiber degenerates
over that divisor. One more remark is in order.
The methods we described and explained apply to the cases in which the 
singularity structure of a threefold (or $n$-fold) can be analyzed in terms
of that of a complex surface, i.e. it is enough to consider a small disc
$D\in \IC$ (or $D\in \IC^{n-2}$) cutting the divisor in the base transversely 
and analyze the 
geometry of the complex surface which is a fibration over $D$. There are cases
which cannot be reduced to that and which are encountered in string theory.
We think that torically one of the signs of such cases is the fact that the 
points irrelevant in the corresponding `top' become relevant in the polyhedron
of the threefold. It would be interesting to find a systematic toric way 
of analyzing such more complicated cases. 

{\it Acknowledgements:} 
We would like to thank P. Candelas, S. Katz and G. Rajesh for useful 
discussions. 
This work was supported by the Austrian ``Fonds zur
F\"orderung der wissenschaftlichen Forschung'' with a Schr\"odinger
fellowship, number J012328-PHY, NSF grant PHY-9511632 and the Robert 
A. Welch Foundation.

\bye